\documentclass[12pt]{iopart}

\usepackage{amssymb, graphicx, braket, cite}

\usepackage[colorlinks=False, linkcolor=black, citecolor=black, urlcolor =black]{hyperref}

\newcommand{\+}{^\dagger}
\newcommand{\trace}{{\rm tr}}
\newcommand{\tO}{\tilde{O}}

\begin{document}

\bibliographystyle{unsrt}

\title{Proposal for measuring out-of-time-ordered correlators at finite temperature with coupled spin chains}

\author{Bhuvanesh Sundar$^{1,2,3}$, Andreas Elben$^{3,4,5}$, Lata Kh Joshi$^{3,4}$, and Torsten V. Zache$^{3,4}$}
\address{$^1$JILA, Department of Physics, University of Colorado, Boulder, CO 80309, USA}
\address{$^2$Center for Theory of Quantum Matter, University of Colorado, Boulder, CO 80309, USA}
\address{$^3$Institute for Quantum Optics and Quantum Information of the Austrian Academy of Sciences, Innsbruck A-6020, Austria}
\ead{bhuvanesh.sundar@colorado.edu}
\address{$^4$Center for Quantum Physics, University of Innsbruck, Innsbruck A-6020, Austria}
\address{$^5$Institute for Quantum Information and Matter and Walter Burke Institute for Theoretical Physics,
California Institute of Technology, Pasadena, CA 91125, USA}

\begin{abstract}
Information scrambling, which is the spread of local information through a system's many-body degrees of freedom, is an intrinsic feature of many-body dynamics. In quantum systems, the out-of-time-ordered correlator (OTOC) quantifies information scrambling. Motivated by experiments that have measured the OTOC at infinite temperature and a theory proposal to measure the OTOC at finite temperature using the thermofield double state, we describe a protocol to measure the OTOC in a finite temperature spin chain that is realized approximately as one half of the ground state of two moderately-sized coupled spin chains. We consider a spin Hamiltonian with particle-hole symmetry, for which we show that the OTOC can be measured without needing sign-reversal of the Hamiltonian. We describe a protocol to mitigate errors in the estimated OTOC, arising from the finite approximation of the system to the thermofield double state. We show that our protocol is also robust to main sources of decoherence in experiments.
\end{abstract}

\section{Introduction}\label{sec: intro}
Present day quantum simulators based on trapped ions~\cite{blatt2012quantum,monroe2021programmable}, ultracold ground state atoms~\cite{gross2017quantum} and Rydberg atoms~\cite{browaeys2020many}, and other platforms~\cite{kjaergaard2020superconducting, vandersypen2005nmr, jones2010quantum, oliveira2011nmr} provide unique opportunities to study the time evolution of quantum many-body systems in a controlled laboratory setting. In atomic physics, analog quantum simulators with ultracold atoms in deep optical lattices realize effective spin models with nearest-neighbor interactions~\cite{gross2017quantum, sun2020realization, brown2019bad, nichols2019spin, mazurenko2017cold}, and simulators with Rydberg tweezer arrays or trapped ion chains realize spin models with longer ranged interactions~\cite{de2019observation, scholl2020programmable, ebadi2020quantum, semeghini2021probing, kokail2019self}. These experiments have several control parameters which allow one to controllably modify, or even completely turn off, these interactions, for example by increasing the lattice depth in optical lattices or detuning the atomic levels in Rydberg systems. This flexibility, and the ability to imprint local operations and make measurements, pave the way to develop quantum algorithms for measuring quantitative probes of the system's non-equilibrium dynamics and thermalization. 

A fundamental question in out-of-equilibrium many-body dynamics is how quantum systems thermalize and scramble information. We describe a protocol to measure a system's finite-temperature out-of-time-ordered correlation (OTOC), which quantifies the nature of information scrambling, and which we will further describe in detail below. Our method is geared towards analog quantum simulators which realize particle-hole symmetric Hamiltonians that satisfy the eigenstate thermalization hypothesis (ETH). The former condition of particle-hole symmetry provides a favorable scenario to simplify the dynamics in our protocol, specifically the backward time evolution in one of the spin chains which is usually challenging to implement, but is not strictly necessary. The latter condition of ETH provides a favorable scenario to approximately prepare the thermofield double (TFD) state~\cite{cottrell2019build, maldacena2018eternal} that is required as the initial state in our protocol, but has not been shown to be a necessary condition for preparing the TFD. As a concrete example, we consider a long-ranged XX model that can be implemented with Rydberg atoms.

Information scrambling, which is the spread of local information through a system's degrees of freedom, is an intrinsic feature of many-body dynamics~\cite{swingle2018unscrambling}. In quantum systems, the effects of scrambling on an initially local operator $V$ are reflected in the Heisenberg time evolution with an underlying Hamiltonian $H$, i.e., $V(t)= \exp(i H t/\hbar)V(0)\exp(-i H t/\hbar)$. One way to quantify this spread is through the square of the commutator of $V(t)$ with another local operator $W(0)$, i.e, $C(t)=\langle|[V(t), W(0)]|^2\rangle$. Here, the expectation value is taken in the thermal state $\rho_\beta= \exp(-\beta H)/Z$, with $Z={\rm tr}(\exp(-\beta H))$ and $\beta$ denotes the inverse temperature, $\beta = 1/(k_BT)$. OTOCs, which we are interested in, are two of the four terms in the expansion of $C(t)$, namely $\braket{W\+(0) V\+(t) W(0) V(t)}$ and $\braket{V\+(t) W\+(0) V(t) W(0)}$.

The OTOC serves as a crucial observable to understand information scrambling in quantum systems~\cite{swingle2018unscrambling,touil2020quantum,iyoda2018scrambling,Lashkari2013, Shenker2014,maldacena2016bound,murthy2019bounds,PhysRevLett.115.131603,Hosur2016a,chen2017out,Hashimoto2017,HydroOTOCEE,GalitskiLiao, MSS_SYK, PhysRevX.7.031047, bohrdt2017scrambling, lin2018out, daug2019detection2, rakovszky2018diffusive, khemani2018operator, klug2018hierarchy}. Generically, OTOCs decay with time until $V(t)$ has reached all the degrees of freedom, in a time scale known as the scrambling time, $t_{\rm scr}$. The scaling of $t_{\rm scr}$ with system size diagnoses how fast or slow the system scrambles information. The decay time and nature of spreading (e.g., diffusive or ballistic) can depend on parameters in the Hamiltonian \cite{MSS_SYK,Stanford2016, PhysRevX.7.031047, GalitskiLiao, belyanski2020minimal, li2020fast} or state properties like the temperature $(T)$ \cite{vijay2018finitetemperature,StanfordYao, Romatschke2021}. In addition to characterizing the nature of scrambling, OTOCs can also be used to diagnose quantum phase transitions~\cite{lewis2020detecting, nie2020experimental, daug2019detection1, daug2020topologically, wei2019dynamical, sun2020out, wang2019probing, heyl2018detecting, shen2017out}. Due to their fundamental importance in information scrambling and other applications listed above, there has been enormous interest in measuring OTOCs in recent years.

There are previous theoretical proposals to experimentally measure OTOCs at infinite temperature~\cite{vermersch2019probing, yoshida2019disentangling, daug2019detection2}. Some of them involve time evolution by reversing the Hamiltonian's sign \cite{swingle2016measuring}, or by controlling the Hamiltonian's sign with an ancillary qubit which acts as a switch \cite{zhu2016measurement, pg2020exponential}. Proposals to measure OTOCs without reversing time evolution involve implementing the SWAP operator between two systems either as an ensemble of random initial states \cite{vermersch2019probing} or in the measurement \cite{yao2016interferometric}, or by making weak measurements \cite{halpern2017jarzynski, dressel2018strengthening}. These methods have been used in experiments to measure the OTOC in nuclear magnetic resonance simulators \cite{li2017measuring, wei2018exploring, nie2019detecting}, ultracold atoms~\cite{pegahan2021energy}, trapped ions \cite{garttner2017measuring, joshi2020quantum, landsman2019verified} and superconducting circuits~\cite{mi2021information, braumuller2021probing, blok2021quantum, wang2021verifying}. 

Refs.~\cite{vermersch2019probing, yoshida2019disentangling} describe how to extend the infinite-temperature methods discussed there to measure the OTOC at finite temperature (referred to as thermal OTOC). Other proposals to measure the thermal OTOC consider two copies of the same system in an entangled initial state prepared at a negative time \cite{lantagne2020diagnosing}, or two copies of the system sampled from a thermal ensemble~\cite{yao2016interferometric}. However, despite these proposals, the OTOC at a finite temperature has not been measured in experiments.

In this work, we describe how to measure the thermal OTOC in analog quantum simulators, without initial state preparation at negative times or the need to evolve backward in time for systems with particle-hole symmetry. Our method is inspired by ideas to prepare the TFD for systems that satisfy ETH~\cite{cottrell2019build, maldacena2018eternal}, and measure the thermal OTOC from the TFD~\cite{lantagne2020diagnosing, maldacena2018eternal}. Our main contribution is to demonstrate a feasible method to measure a thermal OTOC in current quantum simulators. We illustrate this by using a concrete example, a 1D spin model with long-ranged XX interactions, and showing that one can prepare the thermofield double state with reasonable fidelity at modest system sizes, and measure a thermal OTOC by quenching this state. In this context, we propose a method that obtains a good estimate for the OTOC even in the presence of limitations due to dissipative dynamics and systematic limitations of our protocol. We note that the Hamiltonian we consider was recently realized in experiment~\cite{de2019observation}.

This article is organized as follows. In Section~\ref{sec: otoc from tfd}, we review the TFD state and show how it can be used to measure thermal OTOCs. In Section~\ref{sec: approximating tfd}, we numerically show how to approximately access the TFD for a 1D spin Hamiltonian as the ground state of a local parent Hamiltonian on two coupled spin chains. In Section~\ref{sec: otoc from g}, we numerically show that an approximation to the OTOC can be measured from the ground state described in Section~\ref{sec: approximating tfd}, and describe a correction protocol to mitigate the error in this approximation. In Section~\ref{sec: robustness to error}, we numerically show that our correction protocol also mitigates errors that arise from decoherence in experiment. In Section~\ref{sec: connections}, we discuss connections between our protocol and previous proposals to measure the OTOC. We summarize in Section~\ref{sec: summary}.

\section{$O_{th}$ from Thermofield Double state}\label{sec: otoc from tfd}
There are different regularized versions of the thermal OTOC in the literature~\cite{lantagne2020diagnosing, yao2016interferometric, maldacena2016bound}. 
Here, we consider the following thermal OTOC between operators $W$ and $V$, 
\begin{equation} \label{eqn: Oth}
O_{\rm th}(\beta, t) = \frac{ \trace\left( e^{-\beta H/2} W\+ V\+(t) W e^{-\beta H/2} V(t) \right)}{ Z }~.
\end{equation}
In \ref{appen:thermalOtoc}, we state the other definitions of the thermal OTOC and explore their temperature dependence for the model considered in this paper.

The TFD state at temperature $T \equiv 1/(k_B\beta)$ is an entangled state on $2n$ qubits (for an $n$-qubit Hamiltonian), defined as
\begin{equation}\label{eqn: tfd}
\ket{{\rm tfd}(\beta)} = \frac{ \sum_E e^{-\beta E/2}\ket{E}\otimes \ket{E^*} }{ \sqrt{Z} }.
\end{equation}
The sum in~(\ref{eqn: tfd}) runs over the eigenstates $\ket{E}$ of $H$, with respective eigenvalues $E$, i.e., $H\ket E=E \ket E$. We denote $H^*$ and $\ket{E^*}$ as the complex conjugates of $H$ and $\ket{E}$, satisfying $H^*\ket {E^*}= E\ket {E^*}$. 
Preparing $\ket{{\rm tfd}(\beta)}$, which is a highly entangled state, in an experiment is non-trivial and is an active area of research. One technique that has been successfully used in the past uses variational quantum circuits~\cite{zhu2020generation, francis2020body}. Here, we show that $O_{\rm th}$ can be measured if one can access $\ket{{\rm tfd}(\beta)}$.

\begin{figure}[t] \centering
\includegraphics[width=0.7\columnwidth]{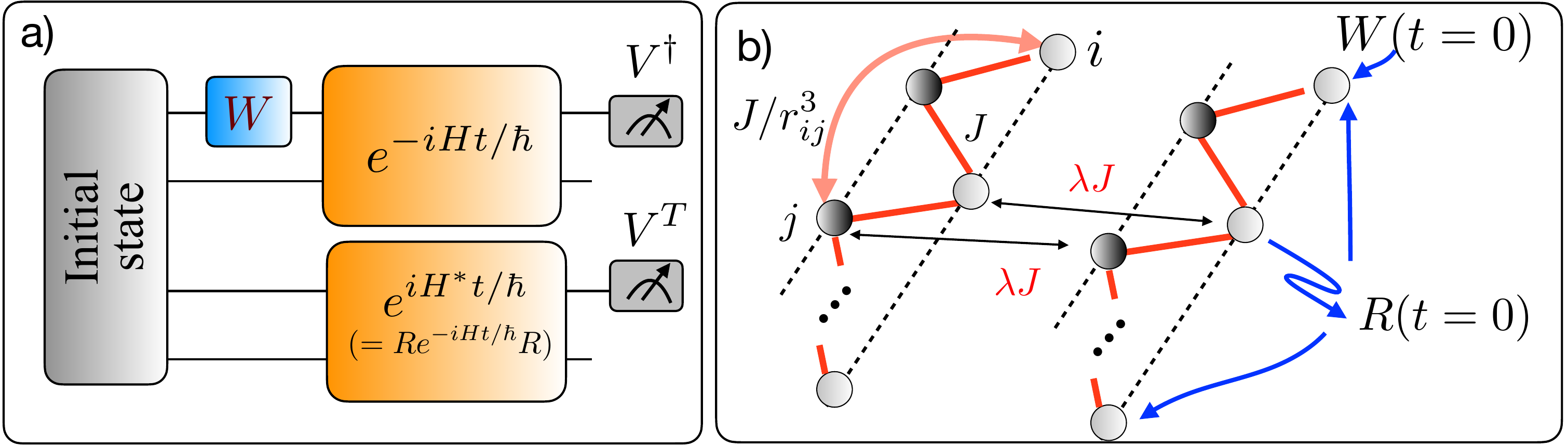}
\caption{(a) Schematic of the protocol to measure the OTOC. We propose to initialize the system in the thermofield double state $\ket{{\rm tfd}(\beta)}$ in the exact protocol, and in the ground state $\ket{g(\lambda)}$ of $H_{\rm parent}$ in the experimentally feasible protocol, which closely approximates $\ket{{\rm tfd}(\beta)}$ up to a symmetry operation (see main text). Then, we apply $W\otimes 1$, and evolve the two halves independently with $H$ and $-H^*$ respectively. For the example we consider [see (b)], $-H^* = RHR$ can be realized by implementing unitary single-qubit gates $R$ before and after time evolution with $H$. Measuring $V\+\otimes V^T$ gives $O_{\rm th}$ in the exact protocol, and $O_g\sim O_{\rm th}$ in the experimental case [see~(\ref{eqn: bound})]. Performing an identical experiment with $W = 1$ gives $N_g$. The corrected estimate for the OTOC is $O_{\rm corr} = O_g/N_g$ [(\ref{eqn: Ocorr})]. (b) Schematic of the physical setup. We consider qubits on a $2\times n$ ladder, with long-range intra-leg interactions given by $J/r_{ij}^3$ [see~(\ref{eqn: H ssh})], and inter-leg interactions given by $\lambda J$ [(\ref{eqn: Hparent ansatz})] which sets the effective temperature. The intra-leg Hamiltonian was recently realized in~\cite{de2019observation}. The initial state is prepared at $t=0$, the inter-leg coupling $\lambda$ is turned off and $W\otimes R$ is applied at $t=0$, and the two legs are evolved independently for $t>0$. }
\label{fig1}
\end{figure}

In an ideal situation where an experiment can access $\ket{{\rm tfd}(\beta)}$, Figure~\ref{fig1}(a) shows a circuit for measuring $O_{\rm th}(\beta, t)$. Realizing the circuit assumes that $W$ is a unitary operator, so that it can be applied in the circuit, and $V$ is Hermitian, so that it can be measured at the end. After preparing the initial state $\ket{{\rm tfd}(\beta)}$, the circuit applies $W\otimes 1$. This step prepares $\ket{\psi_W(\beta, 0)} = (W\otimes1) \ket{{\rm tfd}(\beta)}$. We then evolve the system with $H\otimes 1 - 1\otimes H^*$ for a time $t$, and measure $V\+ \otimes V^T$, leading to,
\begin{equation}\label{eqn: Otfd}
O_{\rm tfd}(\beta,t) = \braket{\psi_W(\beta, t) | V\+ \otimes V^T | \psi_W(\beta,t) }.
\end{equation}
Here, $\ket{\psi_W(\beta, t)} = \exp(-i(H_1-H_2^*)t/\hbar)\ket{\psi_W(\beta, 0)}$, where we denote $H_1 = H\otimes1$ and $H_2 = 1\otimes H$. This protocol requires evolving qubits $[n+1,2n]$ with $-H^*$. We remark that such backward time evolution is a generic feature of protocols that measure the OTOC, except for specific protocols designed to measure the OTOC with only forward time evolution~\cite{vermersch2019probing, joshi2020quantum}.

The measurement $O_{\rm tfd}(\beta, t)$ in~(\ref{eqn: Otfd}) is equal to the thermal OTOC $O_{\rm th}(\beta, t)$ in~(\ref{eqn: Oth}). This can be seen from
\begin{eqnarray}\label{eqn: proof}
O_{\rm tfd}(\beta, t) &= &\bra{ {\rm tfd}(\beta) } (W\+\otimes1) e^{i(H_1-H_2^*)t/\hbar} (V\+ \otimes V^T) e^{-i(H_1-H_2^*)t/\hbar} (W\otimes1) \ket{ {\rm tfd}(\beta) } \nonumber\\
&=& \sum_{EE'} \frac{ e^{-\beta(E+E')/2} }{\trace\ e^{-\beta H}} \braket{ E' | W\+e^{iH t/\hbar}V\+e^{-iH t/\hbar}W | E} \braket{ E'^* | e^{-iH^* t/\hbar}V^T e^{iH^* t/\hbar} | E^*} \nonumber\\
&=& \sum_{EE'} \frac{ e^{-\beta(E+E')/2} }{\trace\ e^{-\beta H}} \braket{ E' | W\+e^{iH t/\hbar}V\+e^{-iH t/\hbar}W | E} \braket{ E | e^{iH t/\hbar}V e^{-iH t/\hbar} | E'} \nonumber\\
&=& \frac{ \trace\left( e^{-\beta H/2} W\+ V\+(t) W e^{-\beta H/2} V(t) \right)}{ \trace\ e^{-\beta H}} \nonumber\\
&=& O_{\rm th}(\beta, t).
\end{eqnarray}
The second line of (\ref{eqn: proof}) is obtained by inserting the definition of $\ket{{\rm tfd}(\beta)}$. Using the general relations $\braket{ E'^* \vert O^* \vert E^*} = \braket{ E' \vert O \vert E}^* = \braket{ E \vert O^\dagger \vert E'}$ for any operator $O$, we obtain the third line of (\ref{eqn: proof}) by setting $O = \exp(i H t/\hbar) V^\dagger \exp(-i H t/\hbar) = V^\dagger(t)$, and noting $(V^\dagger)^* = V^T$. The fourth line is obtained from the identity $\sum_E \exp(-\beta E/2) \ket{E}\bra{E} = \exp(-\beta H/2)$ and the cyclic property of the trace.

A corollary of~(\ref{eqn: proof}) is that setting $W = 1$ yields a measurement that is constant in time. 
This is because $\ket{{\rm tfd}(\beta)}$ is an eigenstate of $H_1-H_2^*$. We denote this special case as $N_{\rm tfd}(\beta)$,
\begin{eqnarray}
N_{\rm tfd}(\beta) &= &\braket{\psi_{W=1}(\beta, t) | V\+ \otimes V^T | \psi_{W=1}(\beta,t) }\nonumber\\
&=& \frac{ \trace\left( e^{-\beta H/2} V\+ e^{-\beta H/2} V \right)}{ \trace\ e^{-\beta H} }.
\end{eqnarray}
Note, $N_{\rm tfd}(\beta) = O_{\rm tfd}(\beta, t=0) \leq 1$. This is unlike some of the other regularizations, where the initial value of the OTOC is 1 [see Appendix~\ref{appen:thermalOtoc}]. We plot $O_{\rm tfd}(\beta, t=0)$ versus temperature in Appendix~\ref{appen:thermalOtoc}.

Next, we present a protocol suitable for analog quantum simulators, inspired by earlier findings in~\cite{maldacena2018eternal, lantagne2020diagnosing, cottrell2019build}, where we approximately realize $\ket{{\rm tfd}(\beta)}$ via the ground state of a local parent Hamiltonian.

\section{Approximating $\ket{{\rm tfd}(\beta)}$ in an atomic quantum simulator}\label{sec: approximating tfd}

For a certain class of Hamiltonians -- those that satisfy the eigenstate thermalization hypothesis (ETH) -- the TFD state has a large overlap with the ground state of a local parent Hamiltonian defined on two appropriately coupled copies of the system~\cite{cottrell2019build}. We consider the Hamiltonian on a single 1D chain to be
\begin{equation}\label{eqn: H ssh}
H = \sum_{i\ {\rm even}} \sum_{j\ {\rm odd}} \frac{J}{r_{ij}^3} \left(\sigma^x_i\sigma^x_j + \sigma^y_i\sigma^y_j \right)
\end{equation}
where $r_{ij}$ is the distance between spins $i$ and $j$, $J$ is the interaction strength between neighboring spins, and the sum runs over $1\leq i,j\leq n$ with $i$ odd and $j$ even. We expect that the long-range interactions will allow the system to thermalize and thus satisfy ETH. Consistent with this, we will show analytical and numerical evidence that the ground state of a local parent Hamiltonian has a large overlap with the TFD state for moderate system sizes $n\leq 11$. We remark that we do not explicitly verify or use the fact that (\ref{eqn: H ssh}) satisfies ETH, but merely use the arguments put forth by earlier works~\cite{cottrell2019build} as a motivation to obtain the TFD as the ground state of a parent Hamiltonian. There are other approaches to prepare the TFD as well, e.g.\ using variational quantum circuits~\cite{zhu2020generation, francis2020body}.

We chose the above Hamiltonian $H$ for two further reasons. First, a recent experiment~\cite{de2019observation} with Rydberg atoms has realized a variant of $H$ where the nearest-neighbor interactions are staggered, i.e.\ a long-range SSH model, by arranging the atoms in a tilted ladder geometry. Realizing the necessary parent Hamiltonian in our protocol [see following sections] could be achieved by placing two such tilted ladders on nearby planes, as shown in Figure~\ref{fig1}(b). The second reason is that the Hamiltonian we consider has particle-hole symmetry, which makes it straightforward to realize the time evolution with $-H^*$ in experiment, which is necessary in our protocol [see Section~\ref{sec: otoc from tfd}]. Particle-hole symmetry means that there exists a unitary $R$ such that $-H^* = R^\dagger H R$, shown as a parenthetical comment in the time evolution operator in Figure~\ref{fig1}(a). For our example, $R = \prod_{n+1\leq k\leq 2n,k\in{\rm even}} \sigma^z_k$ is composed of single-qubit rotations. This is significant because it is easier to realize the evolution operator $\exp(-i(H_1-H_2^*)t/\hbar)$ as $(1\otimes R\+) \exp(-i(H_1+H_2)t/\hbar) (1\otimes R)$ than by physically reversing the sign of the interactions, where we denote $H_1 = H\otimes 1$ and $H_2 = 1\otimes H_2$. The idea to reverse 1D spin Hamiltonians using single-qubit rotations was first put forth in Ref.~\cite{daug2019detection2}. Some particle-hole symmetric models have relatively simple forms for $R$. For example, the nearest-neighbor transverse Ising model has $R_{\rm TIM} = (\prod_{i\ {\rm odd}} \sigma^x_i)(\prod_{i\ {\rm even}} \sigma^z_i)$. However, finding and implementing such $R$ may be infeasible for other cases. We note that particle-hole symmetry is not a requirement; depending on the experimental setup, it may be advantageous to explicitly reverse the Hamiltonian's sign in experiment.

\subsection{Parent Hamiltonian: Special cases}
Let us begin by finding the parent Hamiltonian for $\ket{{\rm tfd}(\beta)}$ for a few special cases.

At $\beta = \infty$, $\ket{{\rm tfd}(\beta)}$ is a product of the ground state of each spin chain. In this case, $\ket{{\rm tfd}(\beta=\infty)}$ is the ground state of
\begin{equation}\label{eqn: H beta=inf}
H_{\beta=\infty} = H_1 + H_2,
\end{equation}
where $H_1 = H \otimes 1$ and $H_2 = 1 \otimes H_2$.

At $\beta = 0$, $\ket{{\rm tfd}(\beta)}$ is a tensor product of EPR pairs, $(\ket{\uparrow_i\uparrow_{i+n}} + \ket{\downarrow_i\downarrow_{i+n}})\sqrt{2}$. Each EPR pair is the ground state of $\sigma^y_i\sigma^y_{i+n} - \sigma^x_i\sigma^x_{i+n}$. Based on this, we can write that $\ket{{\rm tfd}(\beta=0)}$ as the ground state of
\begin{equation}\label{eqn: parent beta=0}
H_{\beta=0} = J\sum_i \sigma^y_i\sigma^y_{i+n} - \sigma^x_i\sigma^x_{i+n}.
\end{equation}

As a final special case, in \ref{appen:n=2}, we analytically derive the parent Hamiltonian for $\ket{{\rm tfd}(\beta)}$ for $n=2$ at arbitrary temperature, finding that $\beta$ is uniquely set by $\lambda$.

We remark that the form of interactions in (\ref{eqn: parent beta=0}) are such that they are ferromagnetic in the $\sigma^x$ direction, and anti-ferromagnetic in the $\sigma^y$ direction for $J>0$ (vice versa for $J<0$), unlike the interactions in $H_1$ and $H_2$ where they are anti-ferromagnetic along both directions. Experimentally, it is however more convenient to realize a system where the interactions have the same sign along all directions. For example, dipole interactions give rise to flip-flip interactions, which yield the same sign for interactions along $x$ and $y$. To overcome this inconvenient scenario, we consider the coupling Hamiltonian
\begin{equation}
H_{12} \equiv S H_{\beta=0} S = J\sum_i \sigma^y_i\sigma^y_{i+n} + \sigma^x_i\sigma^x_{i+n}
\end{equation}
where $S = \prod_{i=1}^n \sigma^y_i$, and note that the ground state $\ket{\phi(\beta=0)}$ of $H_{12}$ is related to $\ket{{\rm tfd}(\beta=0)}$ as $\ket{\phi(\beta)} = S\ket{{\rm tfd}(\beta)}$. We also note that $\ket{\phi(\beta=\infty)} \equiv S\ket{{\rm tfd}(\beta=\infty)}$ is a ground state of $H_1 + H_2$ from (\ref{eqn: H beta=inf}).

\subsection{Parent Hamiltonian: General case}
For intermediate temperature, $0 < T < \infty$ (i.e. $0 < \beta < \infty$), we make an ansatz for the parent Hamiltonian,
\begin{equation}\label{eqn: Hparent ansatz}
H_{\rm parent}(\lambda) = H_1 + H_2 + \lambda J \sum_i (\sigma^y_i\sigma^y_{i+n} + \sigma^x_i\sigma^x_{i+n}).
\end{equation}
We numerically show below that the ground state of this Hamiltonian, $\ket{g(\lambda)}$, has a large fidelity with $\ket{\phi(\beta)} = S\ket{{\rm tfd}(\beta)}$, where the fidelity is defined as
\begin{equation}
F(\beta, \lambda) = \left| \braket{ g(\lambda) \vert \phi(\beta) }\right|^2.
\end{equation}

It is possible to verify whether (\ref{eqn: Hparent ansatz}) is a good ansatz, for each $\beta$, by numerically searching for the value of the inter-chain coupling $\lambda$ that maximizes $F(\beta, \lambda)$. The ansatz is good if $F(\beta, \lambda) \sim 1$. Moreover, we will use $F(\beta, \lambda)$ to put a (loose) bound on the error in the OTOC provided by our protocol. The coupling $\lambda$ may be tuned in experiments by adjusting the distance between the chains, or their orientation relative to the quantization axis, and the value of $\lambda$ sets the inverse temperature $\beta$ realized by the state. Following this intuition, we numerically search for $\beta_0 = {\rm argmax}_\beta F(\beta,\lambda)$, for each $\lambda$.

\subsection{Numerical results for $\beta$} \label{subsec: T vs lambda}

\begin{figure}[t] \centering
\includegraphics[width=0.7\columnwidth]{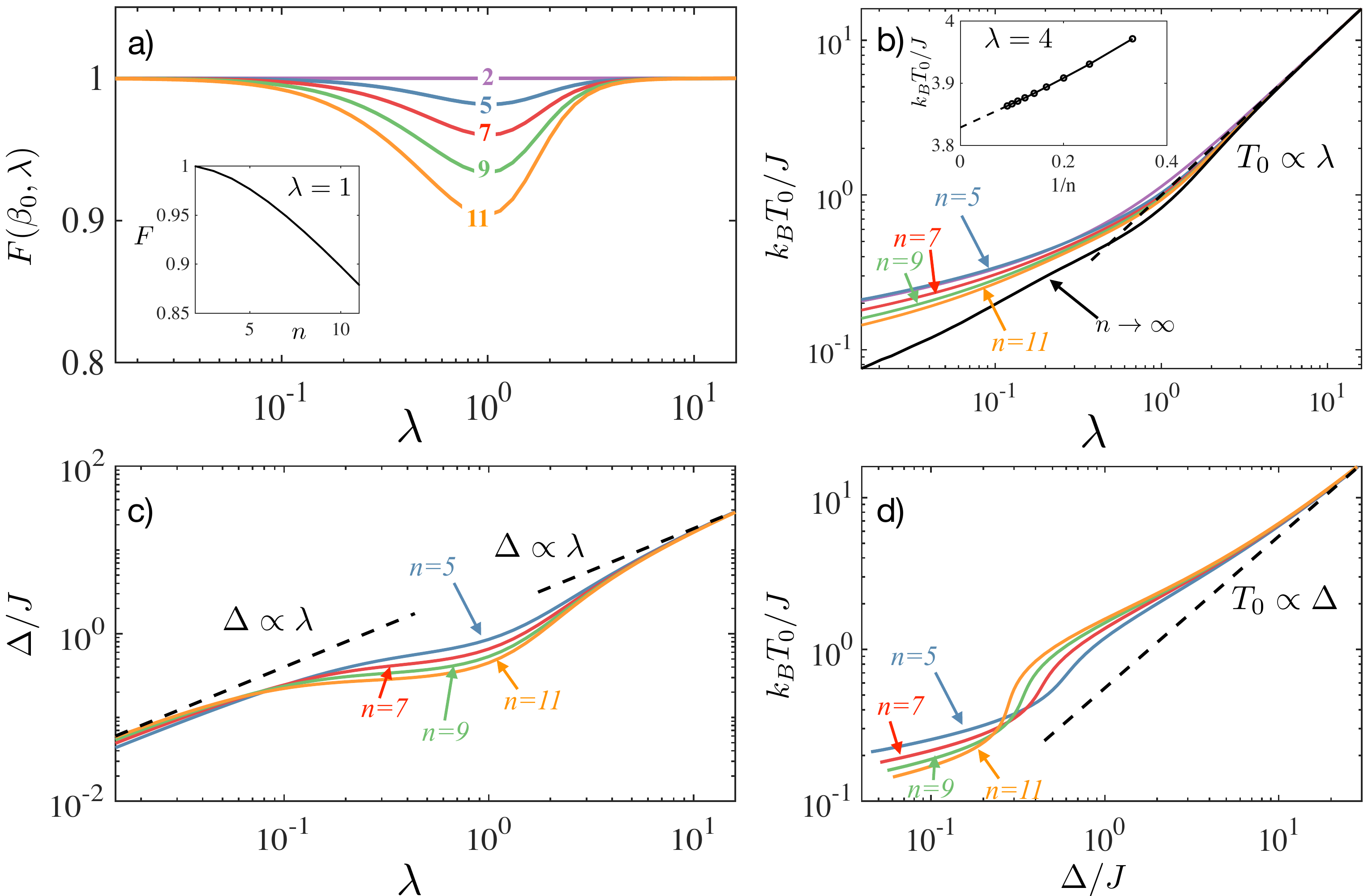}
\caption{(a) Maximum fidelity $F(\beta_0,\lambda) = {\rm max}_\beta F(\beta,\lambda)$ versus inter-leg coupling $\lambda$. Inset shows $F(\beta_0,\lambda)$ versus $n$ at $\lambda=1$. $F(\beta_0,\lambda)\gtrsim0.88$ for $n\leq11$. (b) Effective temperature $T_0$, where the fidelity is maximum, versus inter-leg coupling $\lambda$. 
We extrapolate $T_0$ to $n\rightarrow\infty$ by doing a linear fit of $T_0$ versus $1/n$. Inset shows this extrapolation for $\lambda=4$. (c) The many-body energy gap $\Delta$ of $H_{\rm parent}$ versus $\lambda$. The gap scales linearly with $\lambda$ except in a region around $\lambda=1$. (d) The effective temperature $T_0$ versus the gap $\Delta$.
}
\label{fig2}
\end{figure}

Figure~\ref{fig2}(a) plots the maximum fidelity $F(\beta_0,\lambda) = {\rm max}_\beta F(\beta,\lambda)$ versus $\lambda$, and Figure~\ref{fig2}(b) plots the temperature $T_0=1/(k_B\beta_0)$ where this maximum occurs. The optimum fidelity is always $\gtrsim 88\%$ up to $n=11$, and is smallest around $\lambda=1$ for all $n$. The inset shows the scaling of $F(\beta_0,\lambda)$ with $n$ at $\lambda=1$. The temperature $T_0$ corresponding to the maximum fidelity smoothly increases with $\lambda$ and varies with $n$. We extrapolate $T_0$ to $n=\infty$, by fitting a straight line to $T_0$ versus $1/n$ as shown in the inset of Figure~\ref{fig2}(b). Consistent with perturbation theory~\cite{peschel2011relation}, $T_0$ increases linearly with $\lambda$ for $\lambda\gg1$.

We point out that the fidelity is smallest near $\lambda \sim 1$, and therefore our protocol will perform worse around this region than far away from this region. In Section~\ref{sec: otoc from g}, we will describe a heuristic method to measure the OTOC with reasonable accuracy even when $F(\beta,\lambda)$ deviates from 1.

\subsection{Physical intuition for TFD}
We give several intuitive arguments for why coupling the two chains as in $H_{\rm parent}(\lambda)$ gives the TFD, which could be beneficial to readers from different audiences. All of these arguments argue that after tracing out the degrees of freedom in one chain, say the qubits $n+1 \leq i \leq 2n$, the reduced state $\rho_1$ is a thermal state of $H$, which is consistent with having a TFD.

Our first intuitive argument is as follows. Due to the coupling $\lambda$ between the chains, there is entanglement between the chains, and therefore each chain has a nonzero entanglement entropy. This entanglement entropy, at an intuitive level, makes each chain look like a thermal chain whose temperature is determined by $\lambda$. ETH is sufficient, but not strictly necessary, for this intuition to hold. In systems that satisfy ETH, a subsystem (i.e. one chain in our case)  looks thermal.

Our second argument uses the language of modular Hamiltonians.The modular Hamiltonian of a subsystem is the Hamiltonian for which the subsystem is a thermal state, i.e. the reduced density matrix of the subsystem is $\rho = \exp(-H_{\rm modular})$. Several works~\cite{lauchli2012entanglement, peschel2011relation, kokail2021quantum, pourjafarabadi2021entanglement} have shown that $H_{\rm modular}$ of one chain in the many-body ground state on two coupled chains is proportional to the parent Hamiltonian restricted to that chain, i.e. $H_1$ or $H_2$ in this case. The observation that $H_{\rm modular} \propto H_1,H_2$ remarkably appears to hold for a wide variety of lattice models, and was originally motivated by the Bisognano Wichmann theorem or conformal field theory (CFT)~\cite{bisognano1975duality, bisognano1976duality, cardy2016entanglement, qi2012general}.

Our third argument relies on CFT. In~\cite{qi2012general}, the authors considered two CFTs described by Hamiltonians $H_{1/2}$ and coupled by an interaction $\lambda H_{12}$, similar to our case, and showed that the reduced state is $\rho_1 \propto e^{-H_1/k_BT}$. They showed that the effective temperature $T$ is proportional to the energy gap $\Delta$, and that $\Delta \propto \lambda$. For comparison, this is consistent with the construction of the TFD in \cite{maldacena2018eternal}. We expect the low-energy properties of our model to be well described by a CFT, and therefore the reduced states $\rho_1$ and $\rho_2$ to be thermal states of $H$, which is consistent with having a TFD.

Figure~\ref{fig2}(c) provides evidence for the CFT explanation above, by plotting the gap $\Delta$ versus $\lambda$ and demonstrating that the approximate relation $\Delta \propto \lambda$ also holds in this case. Note that the fidelity decreases in the region where we observe a deviation from the linear scaling $\Delta \propto \lambda$. For comparison, we also plot $T_0$ versus $\Delta$ in Figure~\ref{fig2}(d), and find that $T_0\propto\Delta$ for large $\Delta$, but it deviates from this linear scaling for $\Delta \lesssim 10J$.

\section{Measuring the OTOC from $\ket{g(\lambda)}$}\label{sec: otoc from g}

As we showed above, $\ket{g(\lambda)}$ approximates $\ket{\phi(\beta_0)} = S\ket{{\rm tfd}(\beta_0)}$. The protocol that we described in Sec.~\ref{sec: otoc from tfd} requires the initial state to be $\ket{{\rm tfd}(\beta)} = S^\dagger\ket{\phi(\beta_0)}$, which can be (approximately) prepared easily by applying $S^\dagger$ to the initial state $\ket{\phi(\beta)}$. We will consider $W$ and $V$ as single-qubit Pauli operators for simplicity. For this case, the operation $S^\dagger$ need not be applied at all, because $S$ commutes with the time-evolution operator, $\exp(-it(H_1-H_2^*)/\hbar)$, and with $W$ and $V$.

We denote the measurement $\langle V\+ \otimes V^T\rangle$ made with $\ket{g(\lambda)}$ as $O_g(\lambda, t)$, to distinguish it from the thermal OTOC $O_{\rm th}(\beta, t)$. 
The nonzero infidelity between $\ket{g(\lambda)}$ and $\ket{\phi(\beta_0)}$, for finite $\lambda\neq 0$, yields $O_g(\lambda, t) \neq O_{\rm th}(\beta_0, t)$. Formally, the error is loosely bounded for all times $t$ by [see~\ref{appen:upperbound} for proof]
\begin{eqnarray}\label{eqn: bound}
|O_g(\lambda, t) - O_{\rm th}(\beta_0, t)| &< &2 ||V||^2 D(\ket{g(\lambda)}, \ket{\phi(\beta_0)})\nonumber\\ &= &2 ||V||^2 \sqrt{1-F(\beta_0,\lambda)}
\end{eqnarray}
where $D(\ket{g(\lambda)}, \ket{\phi(\beta_0)}) = \sqrt{1-F(\beta_0,\lambda)}$ is the trace distance between $\ket{g(\lambda)}$ and $\ket{\phi(\beta_0)}$, and $||V||$ is the spectral norm of $V$. In practice however, the error is much smaller at initial times, and grows with time.

The error at all times is significantly reduced by estimating a corrected OTOC given by
\begin{equation}\label{eqn: Ocorr}
O_{\rm corr}(\lambda, t)\equiv O_g(\lambda, t)/N_g(\lambda, t)~,
\end{equation}
where $N_g$ is obtained by performing the measurements with $W=1$ in the prepared ground state $\ket{g(\lambda)}$. This ratio closely approximates $\tO_{\rm th}(\beta_0, t) \equiv O_{\rm th}(\beta_0, t)/|O_{\rm th}(\beta_0,0)|$. The intuitive reason for this is that the nonideality of the initial state, $\ket{g(\lambda)} \neq \ket{\phi(\beta_0)}$, drives similar dynamics in $N_g(\lambda,t)$ and $O_g(\lambda, t)$, and this dynamics is partially canceled in the ratio $O_{\rm corr}(\lambda, t)$. Then, much of the remaining dynamics in $O_{\rm corr}(\lambda, t)$ is only due to the scrambling of $W$. We note that this technique has been successfully applied in earlier works~\cite{swingle2018resilience, zhang2019information, vermersch2019probing, joshi2020quantum} to mitigate errors in measuring the OTOC due to decoherence and noise sources. Other protocols also use qubit teleportation as a means to distinguish decay of OTOCs from genuine scrambling versus decay due to decoherence~\cite{yoshida2019disentangling, landsman2019verified, blok2021quantum, wang2021verifying}.

\subsection{Numerical results for the OTOC}\label{sec: results}
\begin{figure}[t] \centering
\includegraphics[width=0.7\columnwidth]{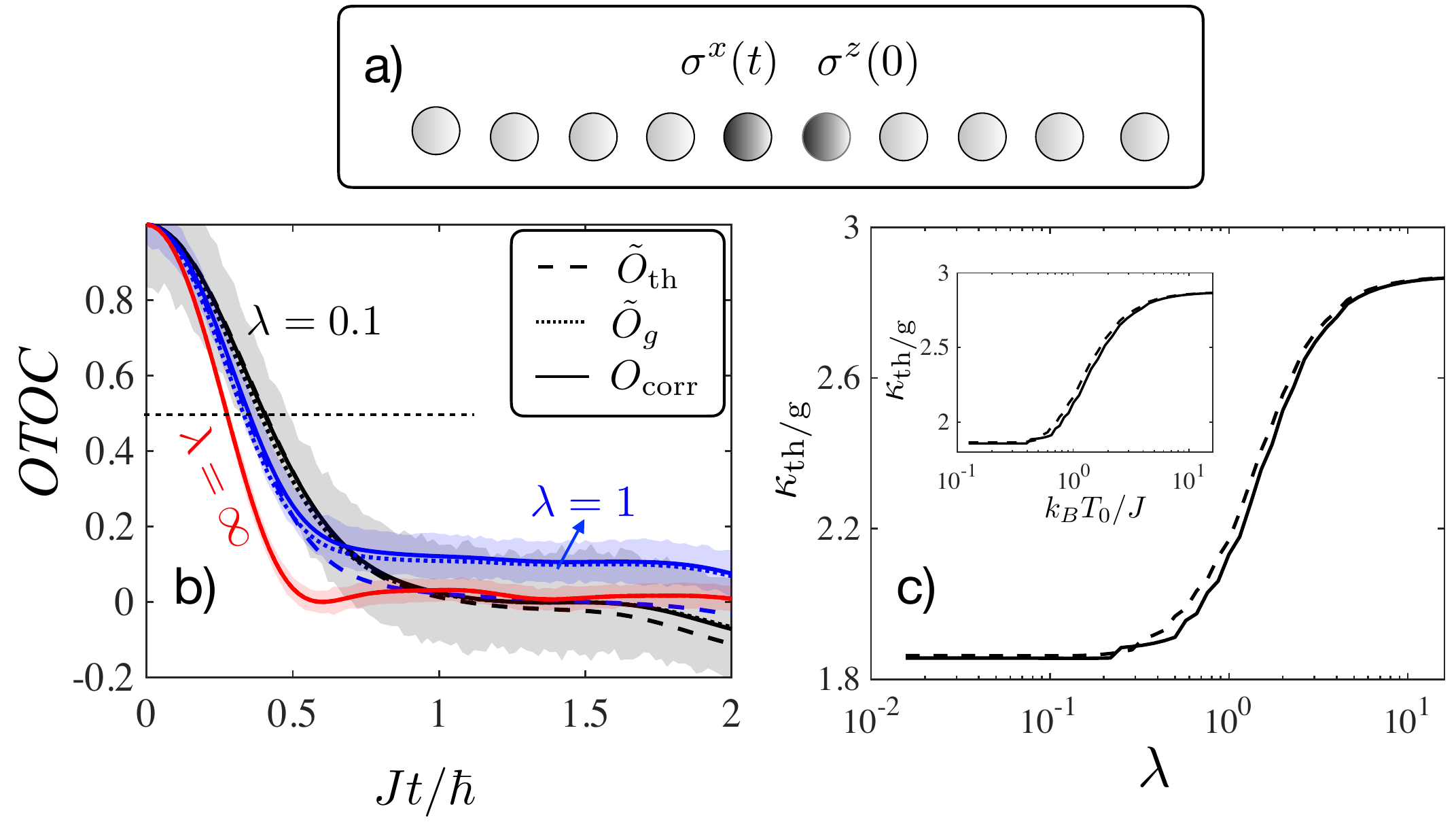}
\caption{(a) A chain of 10 spins for which we calculate the OTOC for $W=\sigma^z_6$ and $V=\sigma^x_5$. (b) OTOCs $\tO_g(\lambda, t) = O_g(\lambda, t)/O_g(\lambda, t=0)$ (dashed), $\tO_{\rm th}(\beta_0, t) = O_{\rm th}(\beta_0, t)/O_{\rm th}(\beta_0, t=0)$ (dotted), and $O_{\rm corr}(\lambda, t) = O_g(\lambda, t)/N_g(\lambda, t)$ (solid), at three different couplings. Shaded areas indicate $1\sigma$ statistical error in $O_{\rm corr}$ from $1000$ measurements of $O_g$ and $N_g$ at each time. The dashed, dotted, and solid curves coincide at $\lambda=\infty$ (red), but differ for the other two couplings. (c) The slope $\kappa = d\tilde{O}/dt$, for $\tilde{O}_{\rm th}$ at $\tO_{\rm th}=0.5$ (dashed), and for $\tO_g$ at $\tO_g=0.5$ (solid). The slope monotonically increases with coupling $\lambda$, and saturates at $\lambda\sim 10$. The inset displays $\kappa$ as function of temperature $T_0=T_0(\lambda)$, with $T_0(\lambda)=1/\beta_0(\lambda)$ determined numerically by maximizing the overlap $F(\beta, \lambda)$(see main text and Figure~\ref{fig2}).
}
\label{fig3}
\end{figure}

\begin{figure}[t] \centering
\includegraphics[width=0.7\columnwidth]{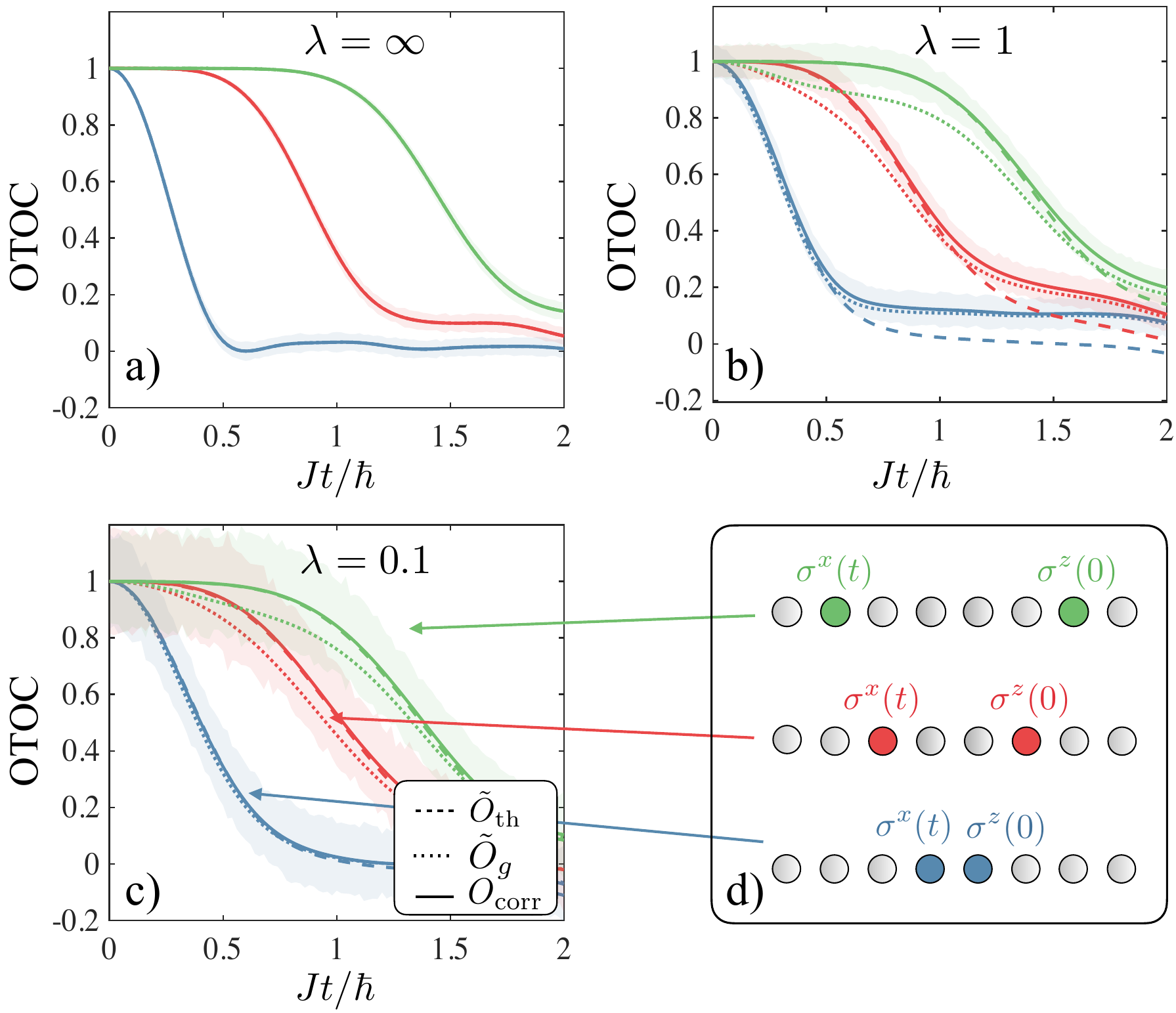}
\caption{(a-c) OTOCs $\tO_g$ (dotted), $\tO_{\rm th}$ (dashed), and $O_{\rm corr}$ (solid) versus time. 
Each panel considers a different coupling $\lambda$. The dotted, dashed, and solid curves coincide at $\lambda=\infty$ (panel a). (d) Choices of $W$ and $V$ in panels (a-c), and their color schemes. Blue lines in (a-c) correspond to $W=\sigma^z_5$ and $V=\sigma^x_4$, red lines to $W=\sigma^z_6$ and $V=\sigma^x_3$, and green lines to $W=\sigma^z_6$ and $V=\sigma^x_2$. $n=8$ in all cases. Shaded areas indicate $1\sigma$ statistical error in $O_{\rm corr}$ from $1000$ measurements of $O_g$ and $N_g$ at each time.}
\label{fig4}
\end{figure}

In this section, we demonstrate that (a) $O_g(\lambda, t)$ approximates $O_{\rm th}(\beta_0, t)$, which is the theoretical OTOC as defined in~(\ref{eqn: Oth}) for inverse temperature $\beta_0$ corresponding to the coupling $\lambda$, (b) the nontrivial temperature-dependence of $O_{\rm th}(\beta_0,t)$ is captured well by choosing $W$ and $V$ as Pauli operators on adjacent spins, $W = \sigma^\mu_i$ and $V = \sigma^\nu_{i-1}$, and (c) errors due to non-ideal fidelity for finite $\lambda\neq0 $ are significantly reduced in $O_{\rm corr}(\lambda, t)$ defined in~(\ref{eqn: Ocorr}).

Figure~\ref{fig3}(b) plots the OTOCs at three different couplings, corresponding to different temperatures [see Figure~\ref{fig2}], choosing $W=\sigma^z_5$, $V=\sigma^x_4$, and $n=8$ as illustrated in Figure~\ref{fig3}(a). Dotted lines correspond to $\tO_{\rm th}(\beta_0, t) \equiv O_{\rm th}(\beta_0, t)/O_{\rm th}(\beta_0,0)$, dashed lines to $\tO_g(\lambda, t) \equiv O_g(\lambda, t)/O_g(\lambda, 0)$, and solid lines to $O_{\rm corr}(\lambda, t) \equiv O_g(\lambda, t)/N_g(\lambda, t)$. The curves start from an initial value of $1$ and decay with time. There is good agreement between them throughout most of the decay. The three curves agree exactly at $\lambda=\infty$ (red curves), but finite size effects and the non-ideality of the initial state for the other two couplings introduce a large disagreement between $\tO_{\rm th}$ and $\tO_g$ at $Jt/\hbar\gtrsim0.5$ and $Jt/\hbar\gtrsim1$. As claimed earlier, the disagreement between $\tO_{\rm th}$ and $O_{\rm corr}$ is smaller than the disagreement between $\tO_{\rm th}$ and $O_g$. Shaded areas plot the $1\sigma$ shot noise from $1000$ measurements. The overall scale of $O_{\rm g}$ decreases as temperature decreases, thereby increasing the shot noise. The shot noise increases with time for the same reason.

The OTOCs' decay rate is different at the three couplings in Figure~\ref{fig3}(b). Figure~\ref{fig3}(c) further illustrates this by plotting the slope $\kappa$ of $\tO_{\rm th/g}$ (solid/dashed) at $\tO_{\rm th/g} = 0.5$. This slope monotonically increases with $\lambda$ and saturates at $\lambda \sim 10$. The inset plots the slope versus $T_0$. The fact that the slope, which is a measure of the rate of the OTOC's decay, depends on the state's temperature is a unique feature which has not been experimentally measured before. A key accomplishment of experimentally implementing our protocol would be measuring this temperature-dependence. 

Figures~\ref{fig4}(a-c) plot the OTOCs for different choices of $W$ and $V$ shown in Figure~\ref{fig4}(d), and for different couplings, corresponding to different temperatures [see Figure~\ref{fig2}], indicated at the top of each panel.

$\tO_{\rm th}$, $\tO_g$, and $O_{\rm corr}$ always agree at $\lambda=\infty$ [Figure~\ref{fig4}(a)], since our protocol is exact in this case. The OTOC stays nearly constant at $\tO_g=1$ for some time before the onset of decay. The time for which it remains nearly constant, $t\sim r/v_B$, is set by the distance $r$ between $W$ and $V$, and the butterfly velocity $v_B$~\cite{roberts2016lieb, roberts2015localized, mezei2017entanglement}. This is because the measurement of $V\+\otimes V^T$ at time $t$ is affected only by the neighborhood where the Heisenberg operator $V\+(t)\otimes V^T(t)$ has sufficient support, and this neighborhood grows linearly with time.

$\tO_{\rm th}$ and $\tO_g$ disagree for finite $\lambda \neq 0$ [Figure~\ref{fig4}(b-c)], and the error is larger when $W$ and $V$ are farther apart from each other. Specifically, $\tO_g$ begins decreasing at $t=0$ while $\tO_{\rm th}$ begins decaying at $t\sim r/v_B$ as explained above. The earlier onset of decay in $\tO_g$ is because $\ket{g(\lambda)}$ evolves with time since it is not an eigenstate of $H_1-H_2^*$. The spins in each leg get correlated with each other during this evolution, and consequently the correlation $V\+\otimes V^T$ across the two legs gets weaker. Remarkably, $O_{\rm corr}$ has a much smaller error with $\tO_{\rm th}$ than the error between $\tO_g$ and $\tO_{\rm th}$.

\section{Robustness to experimental errors}\label{sec: robustness to error}
\begin{figure}[t] \centering
\includegraphics[width=0.7\columnwidth]{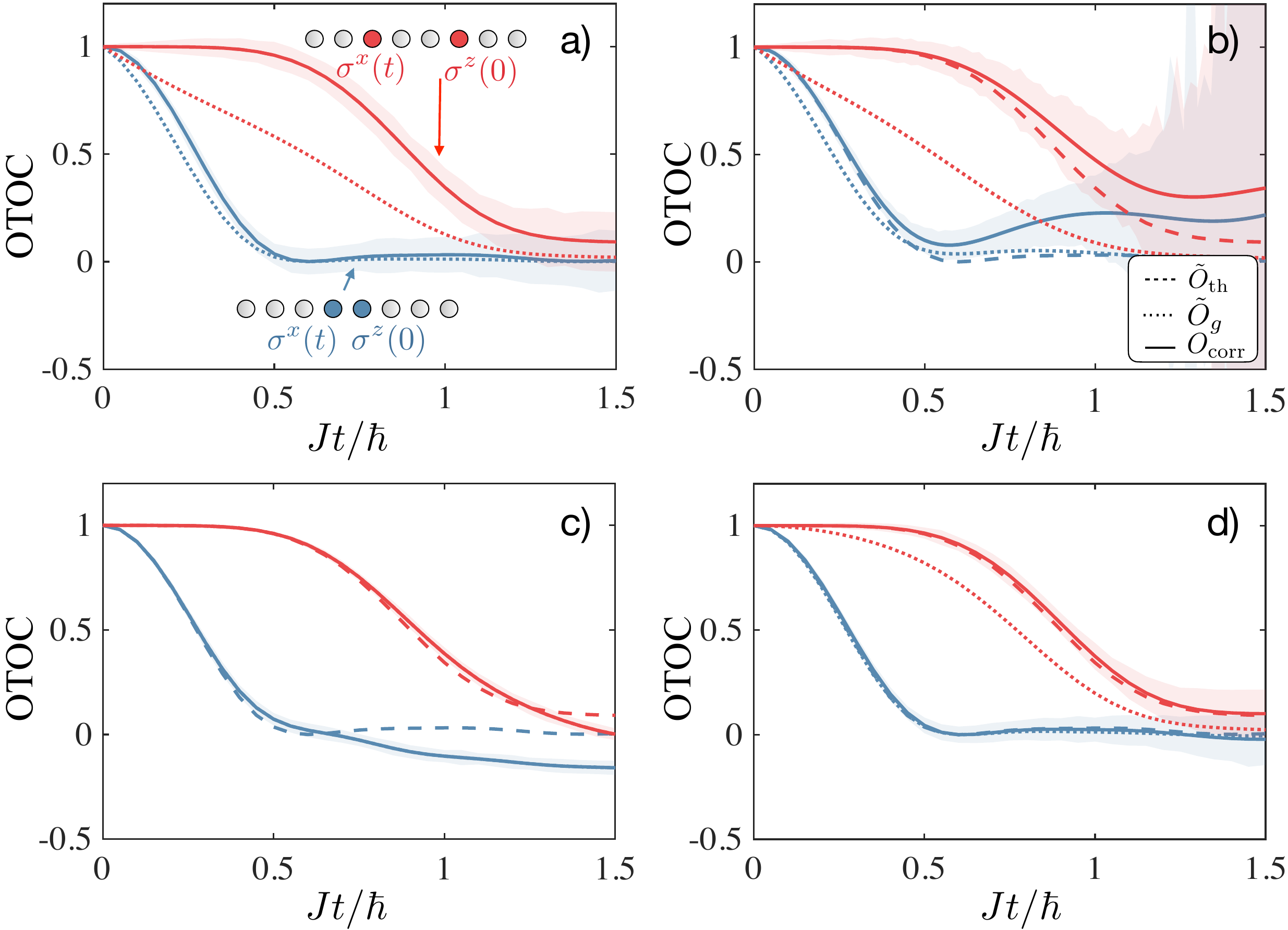}
\caption{OTOCs $\tO_g$ (dotted), $\tO_{\rm th}$ (dashed), and $O_{\rm corr}$ (solid) versus time, in the presence of four error sources. We set $\lambda=\infty$ and $n=8$. (a) considers depolarization with $\gamma=J$. (b) considers local dephasing with $\gamma=J/4$. (c) considers interaction $\epsilon H_{12}$ between the two legs during time evolution, with strength $\epsilon = 0.5$. (d) considers the two legs evolving with unequal Hamiltonians $(1\pm\epsilon)H$, with $\epsilon=0.2$. Blue lines correspond to $W=\sigma^z_5$ and $V=\sigma^x_4$, and red lines to $W=\sigma^z_6$ and $V=\sigma^x_3$, as shown in the two spin chains in (a). Dashed and solid lines completely overlap in (a). Dotted and solid lines completely overlap in (c). Shaded areas indicate $1\sigma$ statistical error in $O_{\rm corr}$ from $1000$ measurements of $O_g$ and $N_g$ at each time.}
\label{fig5}
\end{figure}

\begin{figure}[t] \centering
\includegraphics[width=0.7\columnwidth]{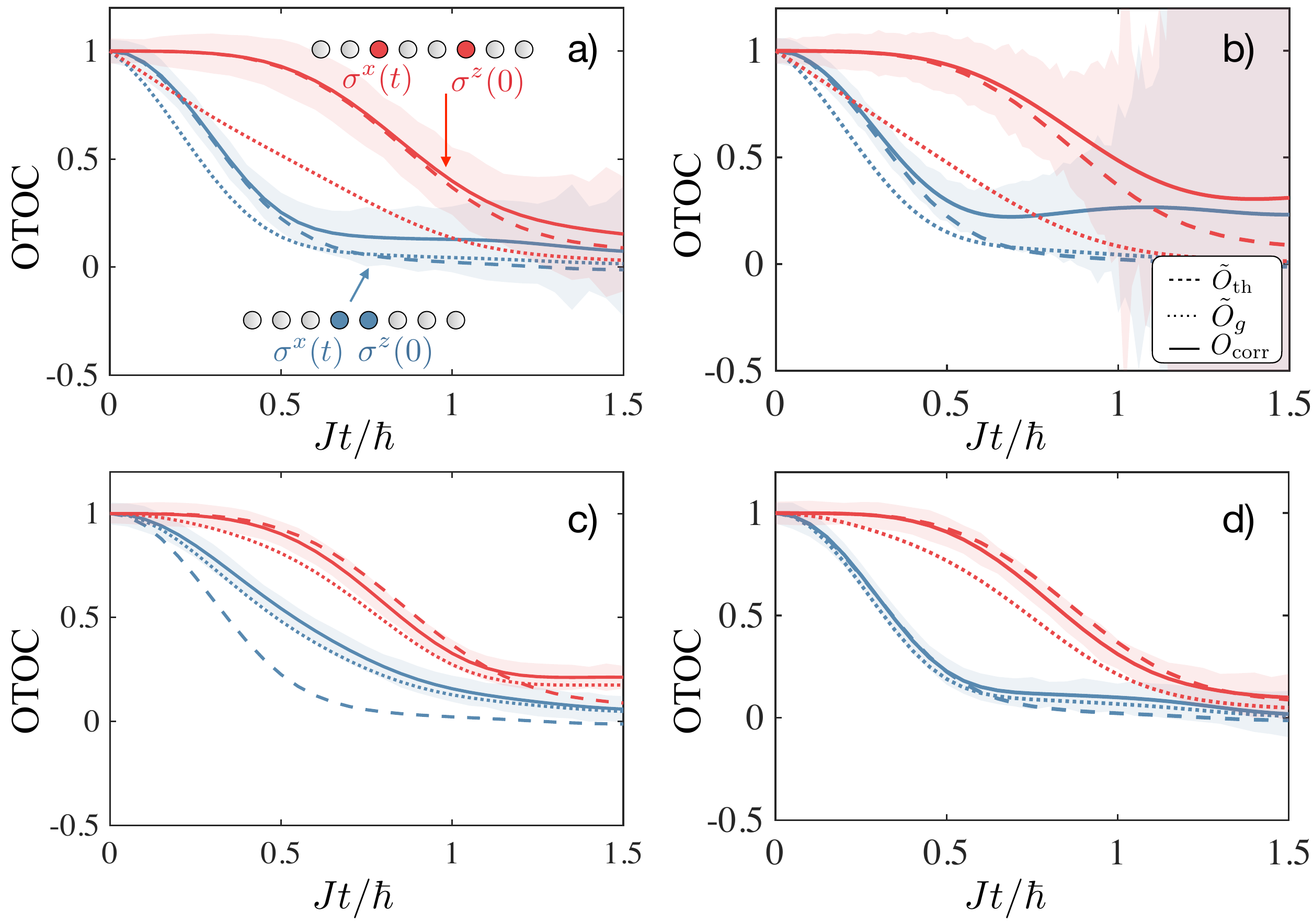}
\caption{OTOCs $\tO_g$ (dotted), $\tO_{\rm th}$ (dashed), and $O_{\rm corr}$ (solid) versus time, in the presence of four error sources. We set $\lambda=1$ and $n=8$. (a) considers depolarization with $\gamma=J$. (b) considers local dephasing with $\gamma=J/4$. (c) considers interaction $\epsilon H_{12}$ between the two legs during time evolution, with strength $\epsilon = 0.5$. (d) considers the two legs evolving with unequal Hamiltonians $(1\pm\epsilon)H$, with $\epsilon=0.2$. Blue lines correspond to $W=\sigma^z_5$ and $V=\sigma^x_4$, and red lines to $W=\sigma^z_6$ and $V=\sigma^x_3$, as shown in the two spin chains in (a). Shaded areas indicate $1\sigma$ statistical error in $O_{\rm corr}$ from $1000$ measurements of $O_g$ and $N_g$ at each time.}
\label{fig6}
\end{figure}

Decoherence in the system during time evolution, or other sources of imperfections, can produce errors in $\tO_g$, even in the limits $\lambda=0$ and $\lambda=\infty$ where our protocol is supposed to be exact. We consider a few different error mechanisms in this section. For the Hamiltonian we consider [(\ref{eqn: H ssh})], we find in all cases except local dephasing that $O_{\rm corr}$ removes these errors at least partially and agrees well with $\tO_{\rm th}$. Note that the bound derived for $|O_g-O_{\rm th}|$ [(\ref{eqn: bound})] only considers errors in the initial state, and is not valid for the error mechanisms in this section.

The first source of decoherence we consider is collective dephasing, caused by fluctuations in the global magnetic field. We model collective dephasing with the Lindblad equation
\begin{equation}
\hbar\partial_t \rho = -i[H_1-H_2^*,\rho] - \gamma \left(S^z\rho S^z - \frac{1}{2} \left( (S^z)^2\rho + \rho (S^z)^2\right) \right),
\end{equation}
where we recall $S^z = \sum_{i=1}^{2n} \sigma^z_i$, and $\rho(\lambda, 0) = \ket{g(\lambda)}\bra{g(\lambda)}$ for the normalization $N_g$ and $\rho(\lambda, 0) = W\ket{g(\lambda)}\bra{g(\lambda)}W\+$ for $O_g$. Remarkably, $N_g$ is completely robust to global dephasing, because $\ket{g(\lambda)}$ is an eigenstate of $S^z$ with eigenvalue $0$ for all $\lambda$, and $H_1-H_2^*$ preserves $S^z$. Thus, the system always lies in a decoherence-free subspace and is insensitive to field fluctuations. $O_g$ is also completely robust to global fluctuations for any $W$ that commutes with $S^z$, e.g., $W = \sigma^z_i$. A similar argument in the Heisenberg picture also implies that $O_g$ is completely robust to collective dephasing if $V\+\otimes V^T$ commutes with $S^z$, e.g., $V = \sigma^z_i$.

The second source of decoherence we consider is depolarization, caused by random collective spin flips along all directions. We model depolarization with the Lindblad equation
\begin{equation}
\hbar\partial_t \rho = -i[H_1-H_2^*,\rho] - \gamma \left(\rho - \frac{1}{4^n} \right).
\end{equation}
This equation has the solution
\begin{equation}
\rho(\lambda, t) = e^{-i(H_1-H_2^*)t/\hbar} \rho e^{-i(H_1-H_2^*)t/\hbar} e^{-\gamma t} + (1-e^{-\gamma t})/4^n.
\end{equation}
It is then straightforward to show that
\begin{eqnarray}
&N_g(\lambda,t;\gamma) = e^{-\gamma t}N_g(\lambda,t;\gamma=0),\nonumber\\
&O_g(\lambda, t;\gamma) = e^{-\gamma t}O_g(\lambda, t;\gamma=0).
\end{eqnarray}
Our protocol is completely robust to depolarization as well, because the factor $e^{-\gamma t}$ in $O_g$ and $N_g$ cancel each other to give the correct value for $O_{\rm corr}$.

Figures~\ref{fig5}(a) and~\ref{fig6}(a) illustrate this robustness to depolarization, at $\lambda=\infty$ and $\lambda=1$, and $\gamma=J$ in both cases. At $\lambda=\infty$, $O_{\rm corr} = \tO_{\rm th}$ exactly. At $\lambda=1$, $O_{\rm corr}\neq \tO_{\rm th}$ because of the finite infidelity of $\ket{g(\lambda)}$ with $\ket{\phi(\beta_0)}$, but no additional errors are introduced by depolarization. The only effect of depolarization on $O_{\rm corr}$ in both cases is that the shot noise is larger than the one in the case of $\gamma=0$. This is because the two factors $O_g$ and $N_g$ in the ratio $O_{\rm corr}=O_g/N_g$ are separately measured in experiment, and each factor has a smaller value with depolarization than without. The smaller values lead to a larger shot noise that increases exponentially with $\gamma$.

The third source of decoherence we consider is local dephasing, caused by magnetic field fluctuations on each spin. We model local dephasing with the Lindblad equation
\begin{equation}
\hbar\partial_t \rho = -i[H_1-H_2^*,\rho] - \gamma \sum_{i=1}^{2n} \left(\sigma^z_i\rho\sigma^z_i - \rho \right),
\end{equation}
Figures~\ref{fig5}(b) and~\ref{fig6}(b) plot $\tO_{\rm th}$ (dashed), $\tO_g$ (dotted) and $O_{\rm corr}$ (solid) at $\lambda=\infty$ and $\lambda = 1$, and $\gamma=J/4$ in both cases. We see a large disagreement between $\tO_{\rm th}$ and $\tO_g$, and a smaller but significant disagreement $\tO_{\rm th}$ and $O_{\rm corr}$.

The fourth and fifth error mechanisms we consider are caused by imperfections in the unitary evolution. In Figures~\ref{fig5}(c) and~\ref{fig6}(c), we consider the case that the coupling between the ladder's two legs is not completely turned off during the time evolution, but there is a remnant coupling $\epsilon H_{12}$ with $\epsilon=0.5$. In Figures~\ref{fig5}(d) and~\ref{fig6}(d), we consider the case that the ladder's two legs have different intra-leg interaction strengths, i.e.\ they evolve with $(1-\epsilon)H_1-(1+\epsilon)H_2^*$. We set $\epsilon = 0.2$. At $\lambda=\infty$, $O_{\rm corr}$ agrees well with $\tO_{\rm th}$ for both the error mechanisms. The agreement between them is poorer at $\lambda=1$.

\begin{figure}[t] \centering
\includegraphics[width=0.5\columnwidth]{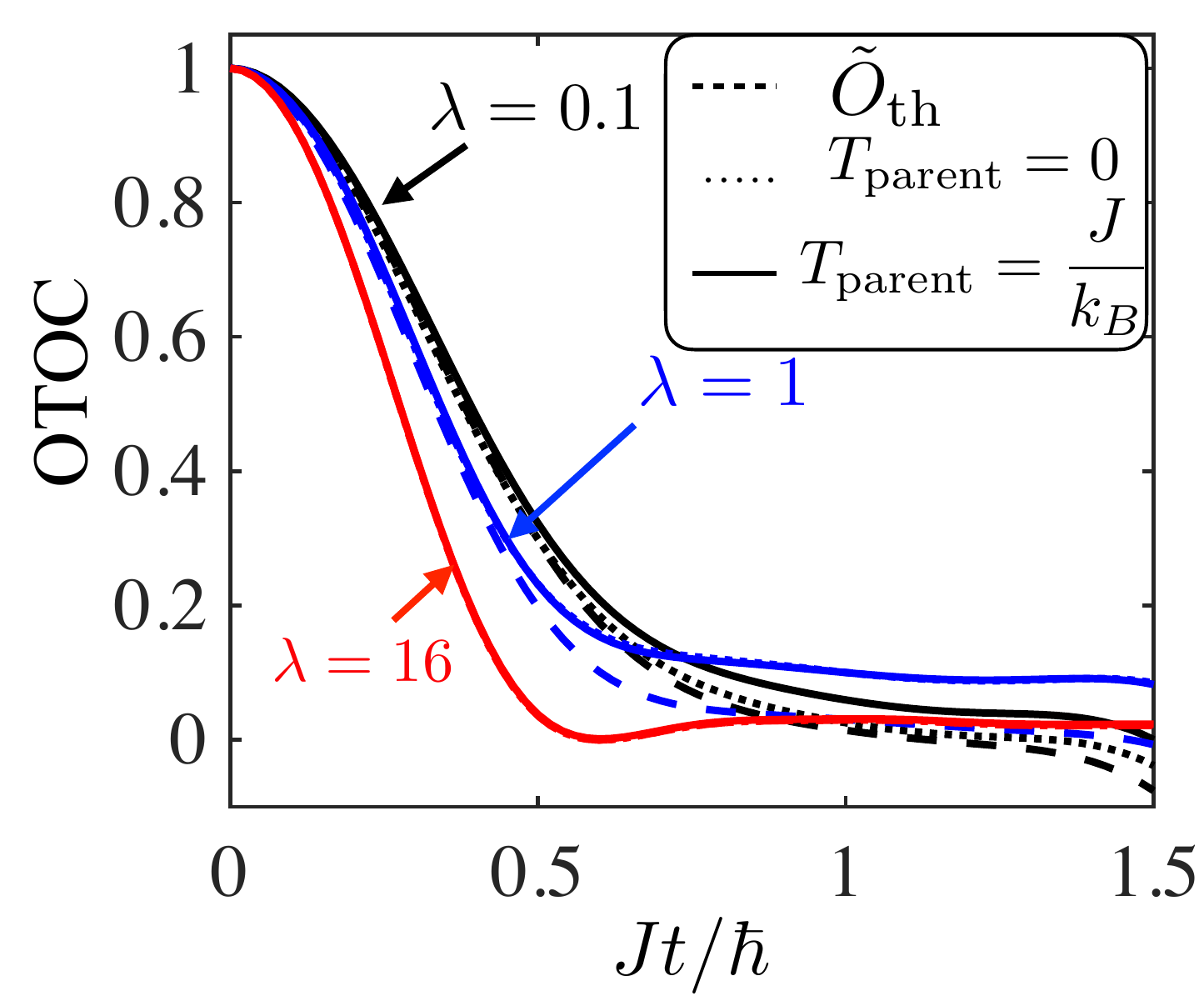}
\caption{OTOCs versus time, when the parent state is at a finite temperature $T_{\rm parent} = J$, i.e. the initial state is $\rho \propto \exp(-H_{\rm parent}(\lambda)/T_{\rm parent}$. Dashed, dotted, and solid lines plot $\tO_{\rm th}$, $O_{\rm corr}$ for $T_{\rm parent}=0$, and $O_{\rm corr}$ for $T_{\rm parent}=J$. The colors correspond to the couplings $\lambda$ indicated in the figure. All three red curves nearly indistinguishable, and the blue dotted and solid lines are also nearly indistinguishable from each other.}
\label{fig7}
\end{figure}

The next error mechanism we consider is due to errors in the initial state, arising from a finite temperature of the parent system. This error is relevant when the parent system is not prepared in the ground state $\ket{g(\lambda)}$, but at a finite temperature $T_{\rm parent}$, i.e. the initial state is $\rho \propto \exp(-H_{\rm parent}(\lambda)/k_BT_{\rm parent})$. Physically, since $H_{\rm parent}(\lambda)$ has a gap $\Delta \propto \lambda$ [see Figure~\ref{fig2}], the initial state would have a large overlap with $\ket{g(\lambda)}$ if $T_{\rm parent} \ll \lambda J$. Moreover, we expect that our correction protocol will mitigate this error, just as it did for the nonzero infidelity of $\ket{g(\lambda)}$ with $\ket{\phi(\beta)}$. We show evidence for this claim in Figure~\ref{fig7}, where we plot the OTOCs for the case the parent system has a temperature $T_{\rm parent} = J/k_B$. We find that the agreement between $O_{\rm corr}$ and $\tO_{\rm th}$ at $T_{\rm parent}=J/k_B$ is comparable to the agreement at $T_{\rm parent}=0$. We further analyse the error $|O_{\rm corr} - \tO_{\rm th}|$ in~\ref{appen:errors}.

In addition to mitigating errors due to imperfections in the initial state ($F(\beta_0,\lambda) < 1$), and due to decoherence during time evolution, $O_{\rm corr}$ also corrects errors in measurement. This is the sixth source of error we consider. If the ideal probability to measure $V\+ \otimes V^T$ as $(\pm 1,\pm 1)$ after time evolution is $P_{\pm 1,\pm 1}$, and the error probability for each measurement is $x$, then the ideal and incorrect expectation values of $V\+ \otimes V^T$ are respectively
\begin{eqnarray}\label{eqn: readout errors}
\langle V\+ \otimes V^T \rangle_{\rm ideal} &= &\sum_{\sigma_1,\sigma_2=\pm1} \sigma_1\sigma_2 P_{\sigma1,\sigma2},\nonumber\\
\langle V\+ \otimes V^T \rangle_{\rm incorrect} &= &\sum_{\sigma_1,\sigma_2=\pm1} \sigma_1\sigma_2 \left( (1-x)^2P_{\sigma_1,\sigma_2} + x(1-x)(P_{\sigma_1,-\sigma_2} + P_{-\sigma_1,\sigma_2}) \right.\nonumber\\
&& \left. + x^2 P_{-\sigma_1,-\sigma_2}\right).
\end{eqnarray}
Simplifying the second line of~(\ref{eqn: readout errors}) gives
\begin{eqnarray}
\langle V\+ \otimes V^T \rangle_{\rm incorrect} = (1-2x)^2 \langle V\+ \otimes V^T \rangle_{\rm ideal}.
\end{eqnarray}
Thus, readout errors rescale the ideal expectation value by $(1-2x)^2$, in both the factors $O_g$ and $N_g$. The ratio $O_{\rm corr}$ completely removes readout errors.

\section{Connections to earlier works}\label{sec: connections}
At infinite temperature, our proposed method is related to two protocols presented earlier in Refs.~~\cite{vermersch2019probing, yoshida2019disentangling}. These have been experimentally implemented to measure the infinite-temperature OTOC in trapped ion quantum simulators~\cite{joshi2020quantum, landsman2019verified} and superconducting circuits~\cite{blok2021quantum, wang2021verifying}.

In Refs.~\cite{vermersch2019probing, joshi2020quantum}, a protocol which employed statistical correlations of randomized measurements to measure the infinite-temperature OTOC was described and realized. Based on correlating separate (sequential) experimental runs, which are initialized in randomized product states and evolved forward in time, it requires only single instances of the $n$-qubit quantum system. The key idea is to rewrite $O_{\rm th}(\beta=0,t)$ as
\begin{equation}
O_{\rm th}(\beta=0,t) = \trace\left( {\rm SWAP}\cdot W\+V\+(t)W \otimes V(t)\right)/2^n.
\end{equation}
with ${\rm SWAP}=\sum_{\vec{x},\vec{x}'}\ket{\vec{x}}\bra{\vec{x}'}\otimes \ket{\vec{x}'}\bra{\vec{x}} $ where the sum runs over all computational basis states $\ket{\vec{x}}$, parametrized by bit strings $\vec{x}=(x_i)_{i=1,\dots n}$ with $x_i=0,1$.
As shown in~\cite{vermersch2019probing}, the crux of the method consists in effectively realizing the \emph{two-copy observable} ${\rm SWAP}/4^n$ as an average of randomized initial product states \emph{on single copies} which can be prepared in separate (sequential) experimental runs,
\begin{equation}
\frac{{\rm SWAP}}{4^n} = \sum_{\vec{x}} (-2)^{-|\vec{x}|} \overline{ (u\ket{\vec{x}}\bra{\vec{x}}u^\dagger) \otimes (u\ket{\vec{0}}\bra{\vec{0}}u^\dagger ) }.
\label{eq:swap_rm}
\end{equation}
Here, $|\vec{x}|\equiv\sum_i x_i$ and the overline denotes the average over local random unitaries $u=\bigotimes_{i=1,\dots n} u_i$, with $u_i$ sampled independently for each spin from unitary 2 designs \cite{vermersch2019probing}. The OTOC is then measured by evolving the system, initialized in the randomized product states $(u\ket{\vec{x}}\bra{\vec{x}}u^\dagger)$ forward in time and measure the operators $W\+V\+(t)W$ and $V(t)$, respectively. We note that in practice, the sum in~(\ref{eq:swap_rm}) can be truncated to obtain a rapidly converging series of modified OTOCs \cite{vermersch2019probing}. 

The ${\rm SWAP}$ operator is not positive semi-definite, i.e.\ it is not a density matrix describing a quantum state. 
The connection to our protocol can however be understood by mathematically applying a transpose operation to qubits $[n+1,2n]$ in~(\ref{eq:swap_rm}). This yields \cite{Watrous2018,Elben2020b}
\begin{equation}
\frac{ \ket{{\rm tfd}(\beta=0)}\bra{{\rm tfd}(\beta=0)} }{2^n} = \sum_{\vec{x}} (-2)^{-|\vec{x}|} \overline{ (u\ket{\vec{x}}\bra{\vec{x}}u\+ )\otimes (u^*\ket{\vec{0}}\bra{\vec{0}} u^T)},
\end{equation}
where $u^T$ and $u^*$ are respectively the transpose and complex conjugate of $u$. The infinite-temperature state $\ket{{\rm tfd}(\beta=0)}\bra{{\rm tfd}(\beta=0)}$ can thus be effectively realized as an average over correlated random initial states, prepared by applying $u$ and $u^*$ to bit strings $\ket{\vec{x}}$ and $\ket{\vec{0}}$ in two separate (sequential) runs of the experiment \cite{Watrous2018,Elben2020b}.

In contrast to the above protocol, we propose in the present work to physically realize $\ket{{\rm tfd}(\beta=0)}\bra{{\rm tfd}(\beta=0)}$ as a pure quantum state on two copies of the systems ($2n$ qubits). As described in Section~\ref{sec: otoc from tfd}, the OTOC is then measured by evolving 
the first copy with $H$ and the second copy with $-H^*$, to measure the operator $W\+V\+(t)W \otimes \left( V(t) \right)^T$, with
\begin{eqnarray}
\left( V(t) \right)^T
&= \exp(-i H^T t) V^T \exp(i H^T t)\nonumber \\ &= \exp(-i H^* t) V^T \exp(i H^* t).
\end{eqnarray}

In summary, the quantum protocol for OTOC measurements presented in this work requires twice the number of qubits as the randomized protocol, and requires fine-tuning to realize time evolution with $-H^* = R\+HR$. It can however be readily extended to arbitrary finite temperatures, by preparing finite temperature TFDs $\ket{{\rm tfd}(\beta)}\bra{{\rm tfd}(\beta)}$, and also requires fewer measurements than the randomized protocol in general.

A previous quantum protocol to measure the infinite-temperature OTOC was implemented in~\cite{landsman2019verified} and~\cite{blok2021quantum, wang2021verifying}, following~\cite{yoshida2019disentangling}. There, the authors prepared $2n$ qubits in $\ket{{\rm tfd}(\beta=0)}$, and one additional independent qubit. Then they evolved the two sets of $n$ qubits independently, similar to our method, and finally measured two qubits in the Bell basis. Because of this measurement basis, they could extract the sum of OTOCs $\sum_{W,V} \trace\left( W\+V\+(t)WV(t) \right)/(16\cdot 2^n)$, where the sum runs over the three Pauli operators and the identity for a spin. 
The sum of these OTOCs is related to the probability that the initially independent qubit teleported from its initial location to a final location. Thus, our protocols are identical at $\beta=0$, except for the presence of the initially independent qubit and choice of the final measurement basis.

We also briefly comment that a previous work~\cite{yao2016interferometric} proposed a method to measure a \textit{differently} regularized finite-temperature OTOC [see~\ref{appen:thermalOtoc}]. In this work, the authors sample $2n$ qubits from thermal ensembles and measures overlap between the qubits after coupling to an ancilla and time evolution. Ref.~\cite{vermersch2019probing} proposed a method to measure a symmetrized variant of the finite-temperature OTOC [see~\ref{appen:thermalOtoc}] at high temperature. In this protocol, the finite-temperature correction to the infinite-temperature OTOC is obtained via statistical correlations between randomized measurements with global random unitaries. Both the above protocols do not require sign reversal of the Hamiltonian. It is also possible to measure the thermal OTOC by sampling states from a thermal ensemble, and making weak measurements together with forward and backward time evolution of the system~\cite{dressel2018strengthening}. All these protocols yield an OTOC that is regularized differently from the OTOC we consider in this paper [see Appendix~\ref{appen:thermalOtoc}].

\section{Discussion \& Conclusion} \label{sec: summary}

We described an experimentally feasible protocol to measure a system's finite-temperature out-of-time-ordered correlation, which is a quantitative probe of the nature of information scrambling. Our method utilizes earlier ideas to prepare the TFD ($\ket{{\rm tfd}(\beta)}$)~\cite{cottrell2019build, maldacena2018eternal}, and measure the thermal OTOC from $\ket{{\rm tfd}(\beta)}$~\cite{lantagne2020diagnosing}. It is geared towards analog quantum simulators with a Hamiltonian respecting particle-hole symmetry and satisfying the ETH, but can also be realized in other analog systems as well as digital quantum simulators. As an example, we considered the long-ranged XX model that was recently realized in~\cite{de2019observation}, and our protocol can be extended to other models. We access the finite temperature state as one half of $\ket{{\rm tfd}(\beta)}$ which is related to the ground state of a local Hamiltonian acting on two copies of the system. For the example we considered, the ground state on two copies of the system well-approximated the desired state for moderate system sizes. In a digital quantum simulator, $\ket{{\rm tfd}(\beta)}$ may be prepared variationally for small systems~\cite{zhu2020generation, francis2020body}. The OTOC can be obtained by measuring local correlations between two halves of the system, after applying a local perturbation and evolving the two halves independently. 

Our protocol works accurately at $T=\infty$ and $T=0$, and works reasonably well in the vicinity of these temperatures. The errors in our protocol are largest near $\lambda = 1$. We described a correction procedure to mitigate the errors in our protocol, which arise from non-ideal fidelity of the prepared initial state with the thermofield double state. We found that this correction procedure yields values for OTOCs that agree well with the exact normalized thermal OTOCs at short times, and mitigate errors due to imperfect initial state, and thus allows us to extract useful physics, e.g. the temperature-dependence of the nearest-neighbor OTOCs' decay. Further, we found that the same normalization procedure also mitigated errors due to decoherence at short times. While our protocol captures short-time dynamics of the OTOCs, it does not capture their steady-state behavior.

There are several directions that future works could explore. They could investigate how to improve the preparation of the TFD for measuring thermal OTOCs as well as a variety of other applications, for example by adding more terms to the inter-chain coupling Hamiltonian. Alternatively, one could use variational techniques in various fashions, either directly as a variational circuit to prepare the TFD as in Ref.~\cite{zhu2020generation}, or using variational ansatze to find the parent Hamiltonian, similar to recent works which adopted variational techniques for the opposite problem of finding the modular Hamiltonian~\cite{kokail2021quantum}. It will be useful to develop other correction methods that complement our protocol to distinguish measured decay of the OTOC due to scrambling from measured decay due to decoherence or imperfect initial state. Protocols to measure the other regularizations also need to be developed and realized in experiment.

We discussed connections between our method at $T=\infty$ and methods adopted in earlier experiments~\cite{landsman2019verified, joshi2020quantum, blok2021quantum, wang2021verifying}. Applying our method in finite temperature systems with a maximally scrambling Hamiltonian will provide a rigorous test for analytical predictions about information scrambling in these systems~\cite{maldacena2016bound, murthy2019bounds}. Our method can be straightforwardly applied in systems where the Hamiltonian's sign can be reversed in experiment explicitly, and in systems with a particle-hole symmetric Hamiltonian where explicitly reversing the Hamiltonian's sign will not be necessary. New algorithms are needed for systems where neither of these are true.

Our method goes beyond the paradigm of the infinite temperature OTOC measured in experiments earlier~\cite{landsman2019verified, joshi2020quantum, blok2021quantum, wang2021verifying}, and could lead to the first experimental measurement of the finite-temperature OTOC. The TFD has been of interest in quantum gravity in the context of wormhole teleportation \cite{Maldacena2003, Maldacena2017, Gao2017}. Recently, this teleportation through wormhole has been shown to be thought of as information scrambling in a coupled chain of qubits \cite{schuster2021manybody}. Our work can set a useful benchmark via the measurement of thermal OTOCs for such `quantum gravity in lab' ideas \cite{brown2021quantum, nezami2021quantum, bhattacharyya2021quantum}.

\section*{Acknowledgments}
Work in Innsbruck was supported by the innovation program under the Grant Agreement No. 731473 (FWF QuantERA via QTFLAG I03769), from the Austrian Science Foundation (FWF, P 32597 N) and by the Simons Collaboration on UltraQuantum Matter, which is a grant from the Simons Foundation (651440, P.Z.). A.E. acknowledges funding by the German National Academy of Sciences Leopoldina under the grant number LPDS 2021-02.

We thank Ana Maria Rey and Murray Holland for a careful reading of the manuscript. We thank Norbert Linke, Alaina Green, and Benoit Vermersch for valuable discussions and comments on the manuscript.

\section*{References}
\bibliography{refs}

\begin{appendix}
\section{Different definitions of thermal OTOC}
\label{appen:thermalOtoc}

\begin{figure}[ht]\centering
\includegraphics[width=0.7\columnwidth]{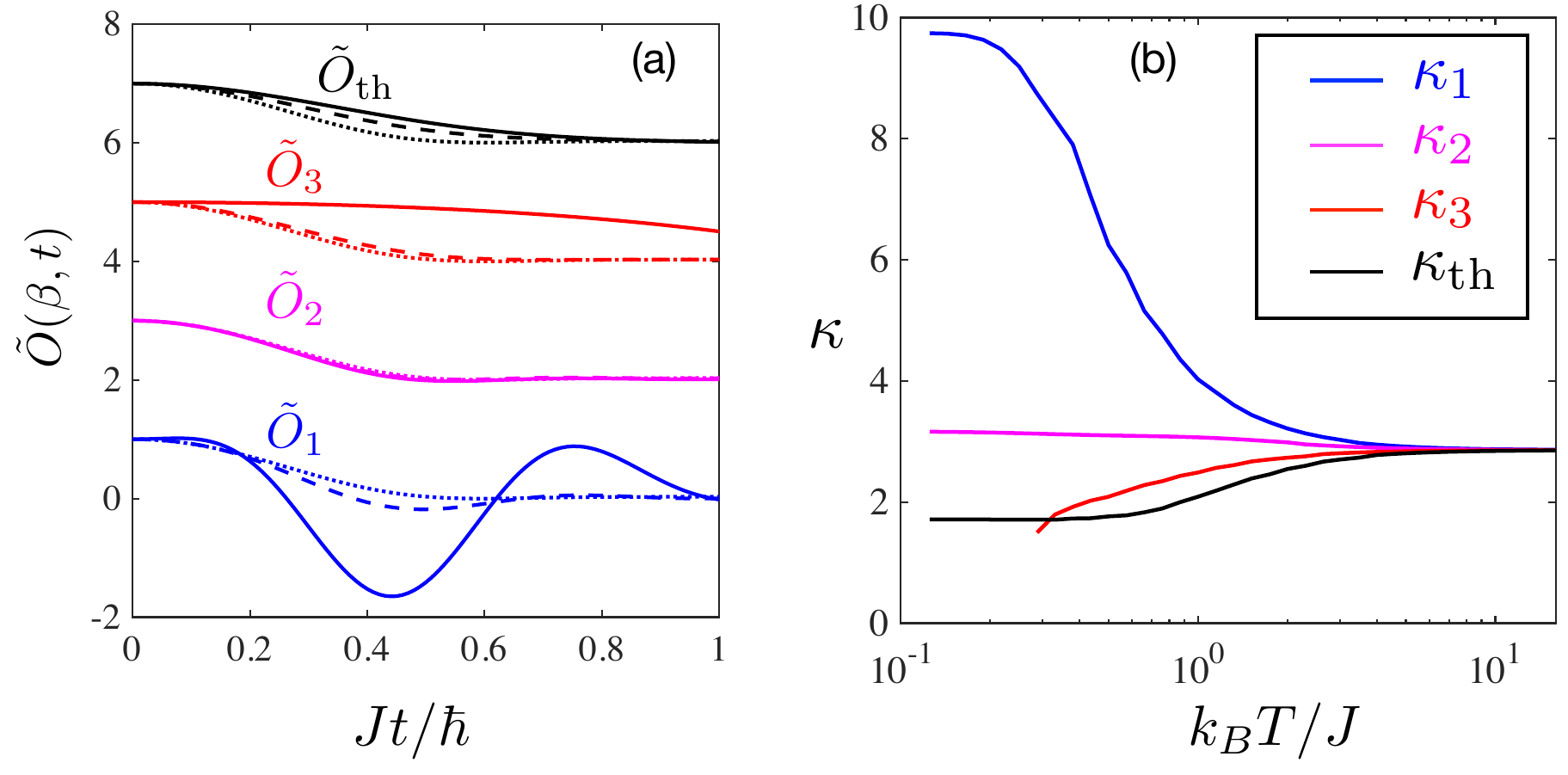}
\caption{(a) Normalized OTOCs $\tO_{1/2/3/\rm th}(\beta, t)$ and (b) their rates of decay, where we define $\tO(t) = O(t)/O(0)$. (a) Blue curves plot $\tO_1(\beta, t)$, magenta curves are $\tO_2(\beta, t)$, red curves are $\tO_3(\beta, t)$, and black curves are $\tO_{\rm th}(\beta, t)$. Solid, dashed, and dotted lines correspond to $k_BT/J = 1/8, 1$, and $8$ respectively. The curves corresponding to $\tO_2$, $\tO_3$, and $\tO_{\rm th}$ are shifted from one another vertically by $2$, $4$, and $6$, respectively for clarity. Each curve's unshifted value at $t=0$ is $1$, due to their normalization. Each curve's unshifted values at $t=0$ is $1$, due to their normalization. (b) The slope of $\tO(\beta, t)$ at $\tO(\beta,t) = 0.5$. The different OTOCs exhibit very different slopes, and even show different trends with temperature. The curves are unshifted in (b).
}
\label{fig: appendix}
\end{figure}

Researchers have used various definitions of the thermal OTOC in the literature. These are:
\begin{eqnarray}\label{eqn: all otocs}
&O_1(\beta, t) = \frac{ \trace\left( y^2 W\+ V\+(t) y^2 W V(t) \right)}{ \mathrm{tr}(e^{-\beta H})}, \nonumber \\
&O_2(\beta, t) = \frac{ \trace\left( y^4 W\+ V\+(t) W V(t) \right)}{ \mathrm{tr}(e^{-\beta H})}, \nonumber \\
&O_3(\beta, t) = \frac{ \trace\left( y W\+ y V\+(t) y W y V(t) \right)}{ \mathrm{tr}(e^{-\beta H})}, \nonumber\\
&O_{\rm th}(\beta, t) = \frac{ \trace\left( y^2 W\+ V\+(t) W y^2 V(t) \right)}{ \mathrm{tr}(e^{-\beta H})},
\end{eqnarray}
where $y = e^{-\beta H/4}$. The definition in the last line is used by us in this paper. 
The above definitions typically show different behaviors from one another for non-maximally-scrambling Hamiltonians at finite temperature. They converge to the same value for $\beta=0$. Refs.~\cite{yao2016interferometric} described a quantum algorithm, based on sampling thermal states, to measure $O_2(\beta, t)$, and~\cite{lantagne2020diagnosing} proposed to measure $O_{\rm th}(\beta, t)$ and $O_1(\beta, t)$ using the TFD. Algorithms to experimentally measure $O_3(\beta, t)$ at arbitrary temperature have not been developed. Here, we investigate their behaviors for the Hamiltonian in~(\ref{eqn: H ssh}).

Figure~\ref{fig: appendix}(a) plots $\tO_{1/2/3/\rm th}(\beta, t)$ for $W=\sigma^z_6$ and $V=\sigma^x_5$ as in the main text, at different temperatures, where we define $\tO(t) = O(t)/O(0)$. All the four kinds of OTOC decay initially with time. But their decay rates are very different from one another. Further, the decay's temperature-dependence is different for the four definitions. While the decay gets steeper as temperature increases for $O_{\rm th}$ and $O_3$, it gets less steep for $O_1$ and is nearly temperature-independent for $O_2$. As before, we compute and plot the curves' slopes at $\tO=1/2$ in Figure~\ref{fig: appendix}(b). The slope $\kappa$ increases with temperature for $O_{\rm th}$ (Figure~\ref{fig2} and black curve in Figure~\ref{fig: appendix}(b)) and $O_3$ (red curve), decreases with temperature for $O_1$ (blue curve), and is nearly constant for $O_2$ (pink curve).

\begin{figure}[ht]\centering
\includegraphics[width=0.45\columnwidth]{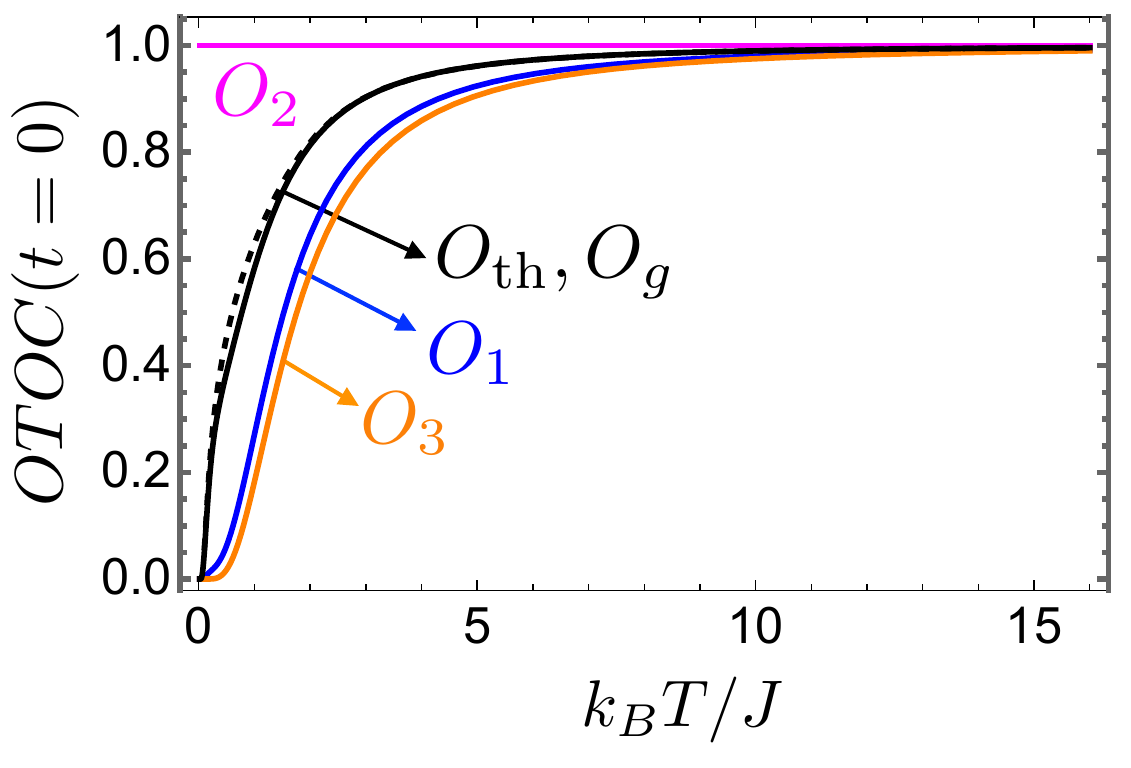}
\caption{Initial (t=0) values of the unnormalized OTOCs defined in Eq.~(\ref{eqn: all otocs}) and in the main text, namely $O_1$ (blue), $O_2$ (magenta), $O_3$ (orange), $O_{\rm th}$ (black), and $O_g$ (black dashed) versus temperature. The unnormalized $O_2(\beta,t=0)$ is always $1$, while the other four increase with temperature. For this system size ($n=10$), $O_{\rm th}(t=0)$ is very close in magnitude to $O_g(t=0)$.
}
\label{fig: vs temp}
\end{figure}

We gain an understanding of the magnitude of the unnormalized OTOCs from Figure~\ref{fig: vs temp}, which plots $O_{1/2/3/\rm th/g}(\beta, t=0)$ versus $k_BT/J$. For $O_g$, we choose the inter-chain coupling $\lambda$ which gives maximum overlap between $\ket{g(\lambda)}$ and $\ket{{\rm tfd}(\beta)}$, as in the main text. The magnitude of the OTOC is $O(\beta, t) = \tO(\beta, t) O(\beta,t=0)$. We find that $O_2(\beta,t=0) = 1$. This is because $W\+V\+WV = 1$ and $\trace(y^4)/\trace(e^{-\beta H}) = 1$. The other four are less than $1$ for all finite $T$, and increase with temperature.

Why these different OTOCs, all of which have been used in the literature, have different values and decay at different rates, and what aspects of information scrambling they capture or miss, are intriguing questions for future exploration.

\section{TFD for $n=2$}
\label{appen:n=2}

Here, we show that $\ket{g(\lambda)} = \ket{\phi(\beta_0)}$ for $n=2$. The ground state of $H_{\rm parent}(\lambda)$ is
\begin{eqnarray}
\ket{g(\lambda)} = &- \frac{\lambda}{2\sqrt{1+\lambda^2}} \left( \ket{\uparrow\uparrow}\ket{\downarrow\downarrow} + \ket{\downarrow\downarrow}\ket{\uparrow\uparrow} \right) \nonumber\\
& - \frac{1}{2\sqrt{1+\lambda^2}} \left( \ket{\uparrow\downarrow}\ket{\uparrow\downarrow} + \ket{\downarrow\uparrow}\ket{\downarrow\uparrow} \right) \nonumber\\
& + \frac{1}{2} \left( \ket{\uparrow\downarrow}\ket{\downarrow\uparrow} + \ket{\downarrow\uparrow}\ket{\uparrow\downarrow} \right).
\end{eqnarray}
Similarly,
\begin{eqnarray}
\ket{\phi(\beta)} = &- \frac{1}{2\cosh\beta} \left( \ket{\uparrow\uparrow}\ket{\downarrow\downarrow} + \ket{\downarrow\downarrow}\ket{\uparrow\uparrow} \right) \nonumber\\
& - \frac{\tanh\beta}{2} \left( \ket{\uparrow\downarrow}\ket{\uparrow\downarrow} + \ket{\downarrow\uparrow}\ket{\downarrow\uparrow} \right) \nonumber\\
& + \frac{1}{2} \left( \ket{\uparrow\downarrow}\ket{\downarrow\uparrow} + \ket{\downarrow\uparrow}\ket{\uparrow\downarrow} \right).
\end{eqnarray}
A straightforward comparison gives $\ket{g(\lambda)} = \ket{\phi(\beta_0)}$ when $\lambda = 1/\sinh\beta_0$.

\section{Upper bound for errors due to the initial state}
\label{appen:upperbound}
For any two states $\rho$ and $\sigma$, the difference in the expectation value of any operator $O$ can be upper bounded as follows. First, we expand $O$ in its eigen basis $\{ \ket{o_\mu} \}$,
\begin{equation}
\braket{O}_\rho - \braket{O}_\sigma = \sum_\mu o_\mu \braket{o_\mu \vert \rho-\sigma \vert o_\mu}.
\end{equation}
Using the triangle sum rule,
\begin{equation}
\braket{O}_\rho - \braket{O}_\sigma \leq \sum_\mu \left| o_\mu \braket{o_\mu \vert \rho-\sigma \vert o_\mu} \right|.
\end{equation}
Then, using $|o_\mu| \leq ||O||$, with $||O||$ being the spectral norm of $O$, and the definition of the trace distance $D(\rho,\sigma)$, we get
\begin{equation}
\braket{O}_\rho - \braket{O}_\sigma \leq 2||O|| D(\rho,\sigma).
\end{equation}
The operator measured in our protocol is $O = V \otimes V^T$, which has spectral norm $||V||^2$. This proves the first line in~(\ref{eqn: bound}), with $\rho=\ket{g(\lambda)}\bra{g(\lambda)}$ and $\sigma=\ket{\phi(\beta_0)}\bra{\phi(\beta_0)}$. These states are pure states, and a standard property \cite{nielsen2002} for pure states leads to the second line in~(\ref{eqn: bound}). The bound is valid for all times, since the spectral norm of a Heisenberg operator is constant with time evolution.

\section{Errors in our protocol}\label{appen:errors}
\begin{figure}[ht]
\includegraphics[width=0.7\columnwidth]{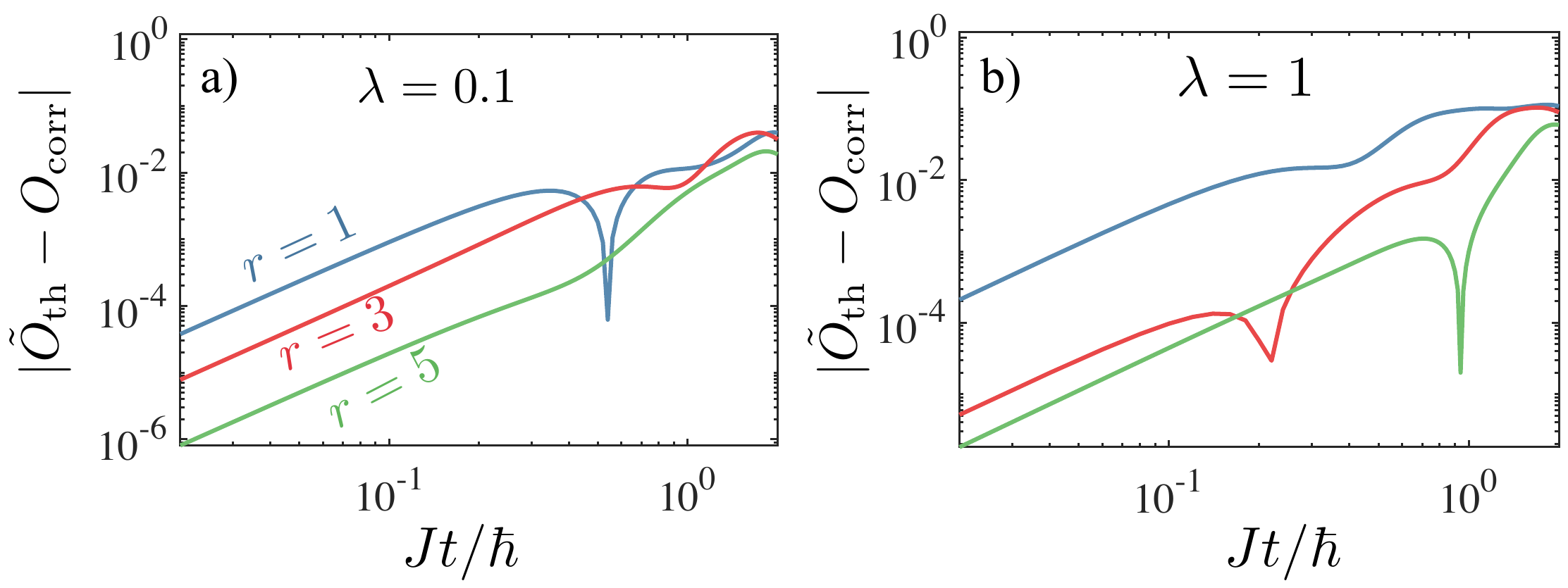}
\caption{The error $|\tilde{O}_{\rm th} - O_{\rm corr}|$ versus $Jt/\hbar$ at (a) $\lambda=0.1$, and (b) $\lambda=1$, for the same operators and distances considered in Fig.~\ref{fig4}. The error $|\tilde{O}_{\rm th} - O_{\rm corr}|$ grows with time, and decreases with $r$ at a fixed time.}
\label{fig: appendix_errors4}
\end{figure}
\begin{figure}[ht]
\includegraphics[width=0.7\columnwidth]{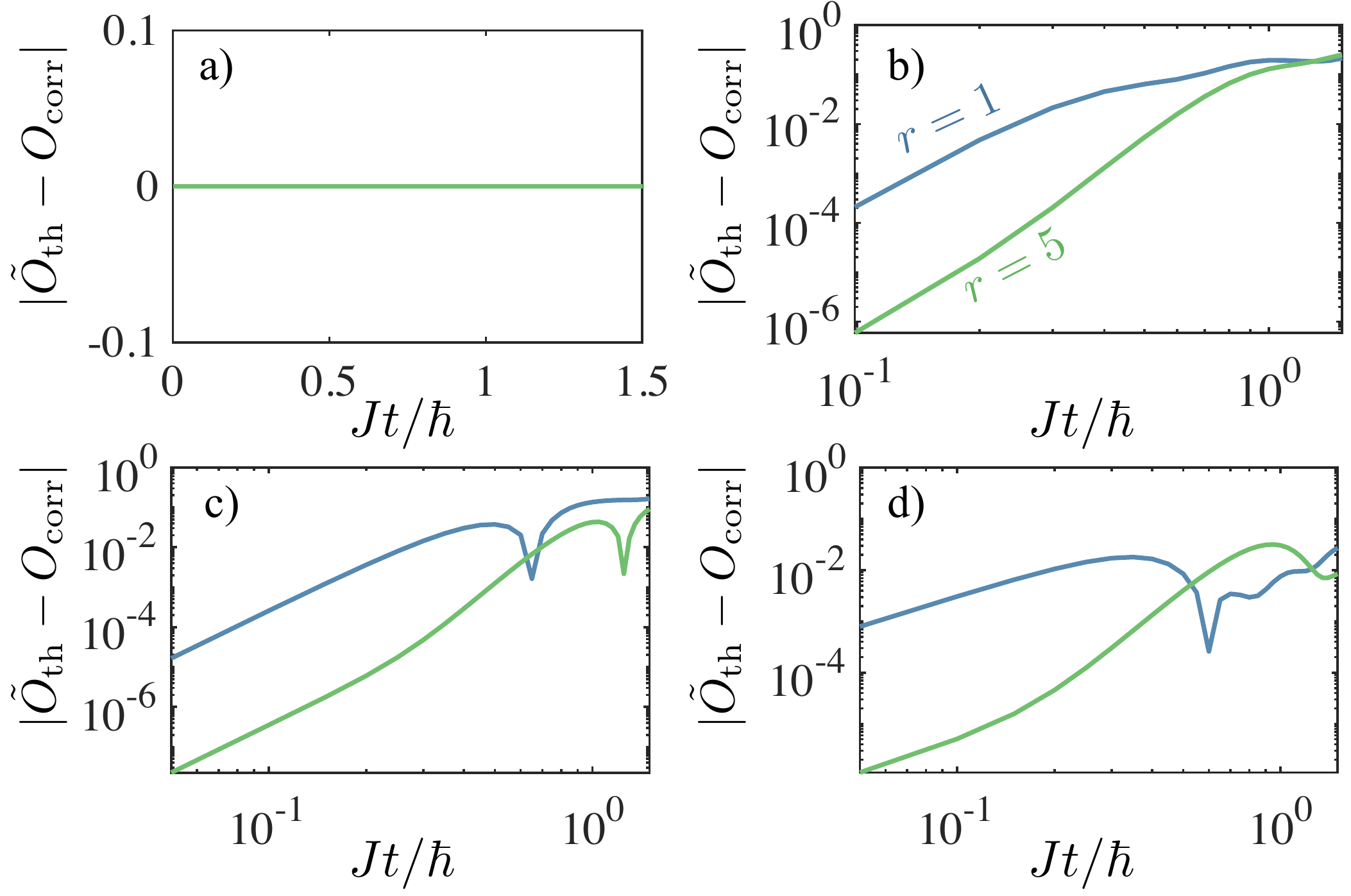}
\caption{The error $|\tilde{O}_{\rm th} - O_{\rm corr}|$ versus $Jt/\hbar$ at $\lambda=\infty$, with the four sources of decoherence considered in Fig.~\ref{fig5}. Each panel plots $|\tilde{O}_{\rm th} - O_{\rm corr}|$ for $r=1$ (blue) and $r=5$ (green), and shows that the error grows with time, and decreases with $r$ at a fixed time.}
\label{fig: appendix_errors5}
\end{figure}
\begin{figure}[ht]
\includegraphics[width=0.7\columnwidth]{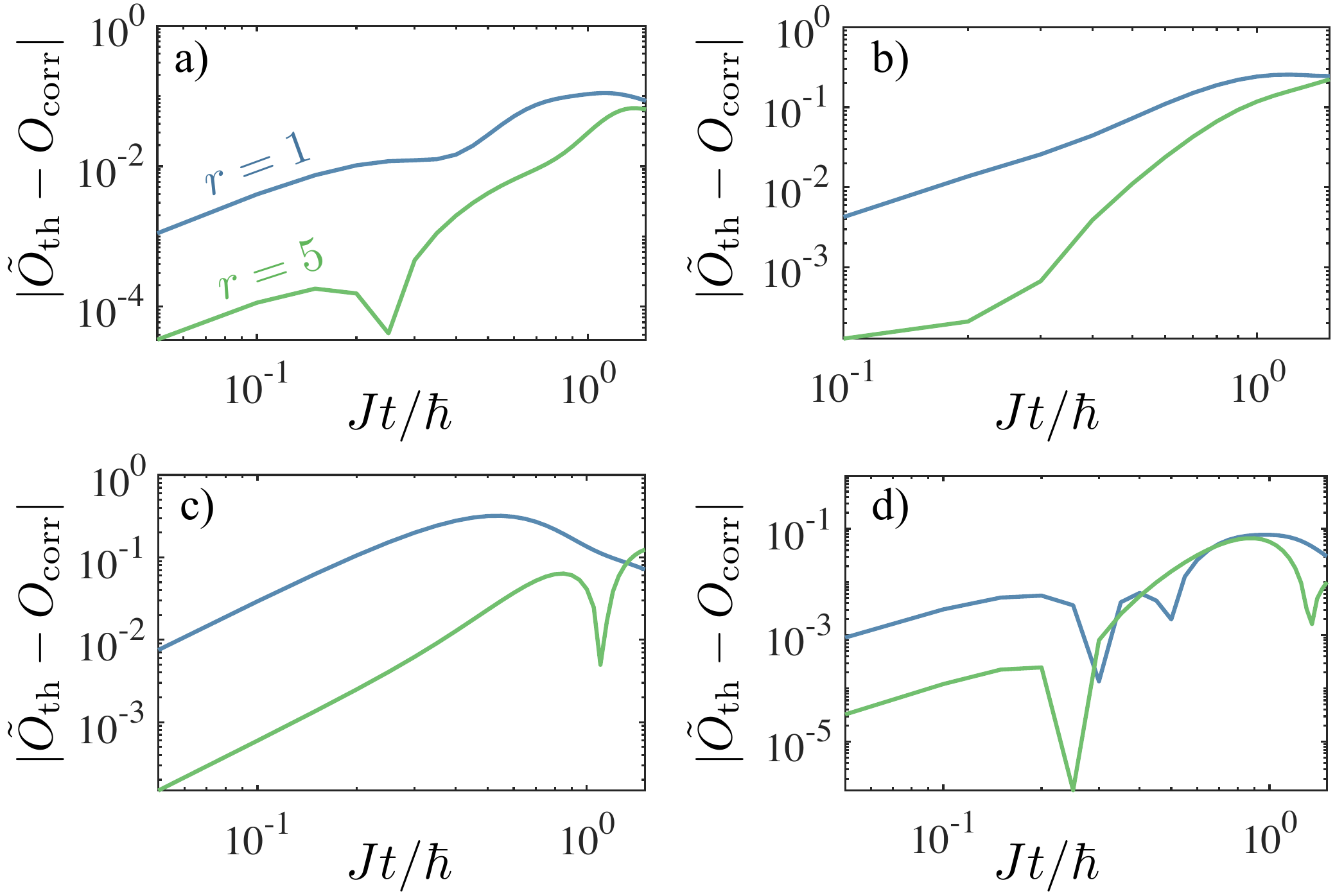}
\caption{The error $|\tilde{O}_{\rm th} - O_{\rm corr}|$ versus $Jt/\hbar$ at $\lambda=1$, with the four sources of decoherence considered in Fig.~\ref{fig6}. Each panel plots $|\tilde{O}_{\rm th} - O_{\rm corr}|$ for $r=1$ (blue) and $r=5$ (green), and shows that the error grows with time, and decreases with $r$ at a fixed time.}
\label{fig: appendix_errors6}
\end{figure}

\begin{figure}[t] \centering
\includegraphics[width=0.7\columnwidth]{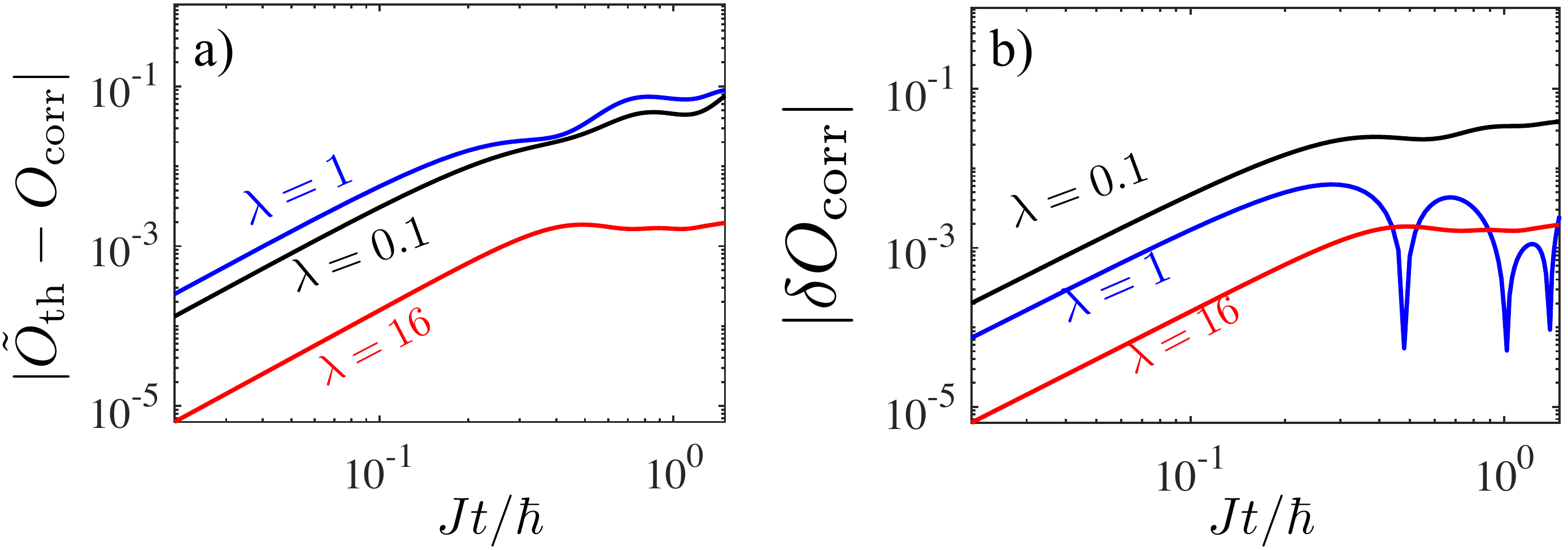}
\caption{(a) The error $|\tilde{O}_{\rm th} - O_{\rm corr}|$ versus $Jt/\hbar$, where $O_{\rm corr} = O(\lambda)/N(\lambda)$ is the corrected OTOC for the case where the parent system's initial state is at a finite temperature, $T_{\rm parent} = J/k_B$. The error at a given time is largest for $\lambda=1$, corresponding to the case that the initial state's fidelity with $\ket{ \phi(\beta_0) }$ is lowest. (b) The difference between $O_{\rm corr}$ at $T_{\rm parent} = 0$ and at $T_{\rm parent} = J/k_B$. The error at a given time is largest for $\lambda=0.1$, corresponding to the case that the parent Hamiltonian's gap $\Delta$ is the smallest.}
\label{fig: appendix_errors7}
\end{figure}

In Figures~\ref{fig3}-\ref{fig7}, we plotted the normalized OTOCs $\tilde{O}_g$ and $\tilde{O}_{\rm th}$, and the corrected OTOC $O_{\rm corr}$. We observed from the figures that $|\tilde{O}_{\rm th} - O_{\rm corr}| < |\tilde{O}_{\rm th} - \tilde{O}_g|$. Here, we analyze the error $|\tilde{O}_{\rm th} - O_{\rm corr}|$ in more detail.

Figure~\ref{fig: appendix_errors4} plots $|\tilde{O}_{\rm th} - O_{\rm corr}|$ versus time, for two couplings $\lambda = 0.1$ and $\lambda = 1$, corresponding to the Figures~\ref{fig4}(b) and (c). We find that the errors increases $\propto t^2$, at short times. We also find that the error decreases with distance at a fixed time, supporting our reason in the main text for correcting $O_g$ with $N_g$.

Figure~\ref{fig: appendix_errors5} plots $|\tilde{O}_{\rm th} - O_{\rm corr}|$ versus time at $\lambda = \infty$, in the presence of the four sources of decoherence considered in Fig.~\ref{fig5}. Figure~\ref{fig: appendix_errors6} plots the errors versus time at $\lambda = 1$, with the same sources of decoherence. Similar to Fig.~\ref{fig: appendix_errors4}, we find in both cases that the error increases with time, and decreases with distance.

Figure~\ref{fig: appendix_errors7} plots the error in the OTOC due to a finite temperature parent system, in two ways. In panel (a), we consider that the parent system is at a temperature $T_{\rm parent} = J/k_B$, and plot the difference between the corrected OTOC and $\tO_{\rm th}$. Among the three coupling strengths $\lambda$, this error is largest at $\lambda = 1$, mainly because the error is largest here even if the parent system is at zero temperature. In panel (b), we again consider that the parent system is at a temperature $T_{\rm parent} = J/k_B$, but plot the difference between the OTOC measured in this state and the OTOC measured from the zero-temperature parent state $\ket{g(\lambda)}$. Here, the error is largest when $\lambda = 0.1$, since the energy gap is the smallest for this case.

\end{appendix}

\end{document}